\documentclass[twocolumn]{aastex631}

\let\tablenum\relax
\usepackage[separate-uncertainty=true, range-phrase=-, range-units=single,
angle-symbol-over-decimal]{siunitx}
\usepackage{amsmath}
\usepackage{graphicx}
\usepackage{chemformula}

\received{}

\shorttitle{}
\shortauthors{}

\DeclareSIUnit{\mag}{mag}
\DeclareSIUnit{\pixel}{pixel}
\DeclareSIUnit{\parsec}{pc}
\DeclareSIUnit{\arcsec}{arcsec}
\DeclareSIUnit{\arcmin}{arcmin}
\DeclareSIUnit{\solarlum}{\mbox{$L_\odot$}}
\DeclareSIUnit{\solarmass}{\mbox{$M_\odot$}}
\DeclareSIUnit{\year}{yr}
\DeclareSIUnit{\deg}{deg}
\DeclareMathOperator{\arcsinh}{arcsinh}
\DeclareSIUnit{\adu}{ADU}
\DeclareSIUnit{\erg}{erg}

\newcommand{\ha}{H$\alpha$}
\newcommand{\hb}{H$\beta$}
\newcommand{\oiii}{[\ion{O}{3}]}
\newcommand{\nii}{[\ion{N}{2}]}

\begin{document}

\title{A Deep Census of Outlying Star Formation in the M101 Group}
\author{Ray Garner, III}
\affiliation{Department of Astronomy, Case Western Reserve University, 10900 Euclid Avenue, Cleveland, OH 44106, USA}

\author{J. Christopher Mihos}
\affiliation{Department of Astronomy, Case Western Reserve University, 10900 Euclid Avenue, Cleveland, OH 44106, USA}

\author{Paul Harding}
\affiliation{Department of Astronomy, Case Western Reserve University, 10900 Euclid Avenue, Cleveland, OH 44106, USA}

\author{Aaron E. Watkins}
\affiliation{Astrophysics Research Institute, Liverpool John Moores University, Liverpool, L3 5RF, UK}
\affiliation{Centre for Astrophysics Research, School of Physics, Astronomy \& Mathematics, University of Hertfordshire, Hatfield, AL10 9A, UK}

\begin{abstract}
We present deep, narrowband imaging of the nearby spiral galaxy M101 and its group environment to search for star-forming dwarf galaxies and outlying \ion{H}{2} regions. Using the Burrell Schmidt telescope, we target the brightest emission lines of star-forming regions, H$\alpha$, H$\beta$, and [\ion{O}{3}], to detect potential outlying star-forming regions. Our survey covers $\sim$\SI{6}{\square\deg} around M101, and we detect objects in emission down to an H$\alpha$ flux level of \SI{5.7e-17}{\erg\per\second\per\square\centi\metre} (equivalent to a limiting SFR of \SI{1.7e-6}{\solarmass\per\year} at the distance of M101). After careful removal of background contaminants and foreground M stars, we detect 19 objects in emission in all three bands, and 8 objects in emission in H$\alpha$ and [\ion{O}{3}]. We compare the structural and photometric properties of the detected sources to Local Group dwarf galaxies and star-forming galaxies in the 11HUGS and SINGG surveys. We find no large population of outlying \ion{H}{2} regions or undiscovered star-forming dwarfs in the M101 Group, as most sources (\SI{93}{\percent}) are consistent with being M101 outer disk \ion{H}{2} regions. Only two sources were associated with other galaxies: a faint star-forming satellite of the background galaxy NGC~5486, and a faint outlying \ion{H}{2} region near the M101 companion NGC~5474. We also find no narrowband emission associated with recently discovered ultradiffuse galaxies and starless \ion{H}{1} clouds near M101. The lack of any hidden population of low luminosity star-forming dwarfs around M101 suggests a rather shallow faint end slope (as flat as $\alpha \sim -1.0$) for the star-forming luminosity function in the M101 Group. We discuss our results in the context of tidally-triggered star formation models and the interaction history of the M101 Group. 
\end{abstract}

\section{Introduction}

Understanding star formation and the variety of locales in which it takes place is key to understanding galaxy formation and evolution. Star formation is most easily seen and studied in the inner luminous regions of galaxies \citep[e.g.][]{martin2001}, but star formation in low-density environments is less well understood, whether that be in low luminosity dwarf galaxies \citep{kennicutt2008,lee2007,lee2009} or outlying \ion{H}{2} regions \citep{rudolph1996,ferguson1998,lelievre2000,ryan-weber2004,werk2010}. In recent years, progress has been made to observationally explore these environments in ultraviolet and optical light. 

UV investigations of the outer regions of galaxies has been primarily aided by the \emph{Galaxy Evolution Explorer} (\emph{GALEX}) satellite. \emph{GALEX} revealed that over $\sim$\SI{30}{\percent} of spiral galaxies posses UV-extensions of their optical disks, possibly indicating that low-density star formation is not rare \citep{thilker2007}. Moreover, extended-UV (XUV) emission and outyling \ion{H}{2} regions beyond the optical radius of the disk are associated with previous or ongoing galaxy interactions \citep{thilker2007,werk2010}.

In optical light, outlying isolated \ion{H}{2} regions have been detected via narrowband imaging targeting specific emission lines, primarily H$\alpha$, where they appear as emission-line point sources. These regions have been found in environments ranging from galaxy clusters and compact groups to the halos of galaxies \citep[e.g.][]{sakai2002,oliveira2004,ryan-weber2004,walter2006,boquien2007,werk2010,kellar2012,keel2012}. These \ion{H}{2} regions indicate recent formation of OB stars outside of the normal star-forming environment of the inner disk. And unlike the XUV emission of spiral galaxies, these outlying \ion{H}{2} regions can appear well outside twice the canonical $R_{25}$ size of the nearest galaxy \citep{ryan-weber2004,werk2010}. This gives us insight into extreme modes of star formation and may contribute to intragroup and intracluster light (see \citealt{vilchez-gomez1999} for a review).

Possible sites for this low-density star formation are dwarf galaxies. Dwarf galaxies are very common around massive galaxies and in group environments. Star forming dwarf galaxies (SFDGs) are one of the most common types of galaxies in the local universe \citep[e.g.][]{lapparent2003}. They are characterized by low stellar mass, low chemical abundance, high gas content, and high dark matter content (see \citealt{gallagher1984} for a review). They tend to lie in low density environments \citep{weisz2011a,weisz2011b} and are typically of low surface brightness. The low surface density of cold gas in SFDGs is below that at which star formation is truncated (the so-called Kennicutt-Schmidt law, e.g.\ \citealt{schmidt1959,kennicutt1989,kennicutt1998}), but star formation is not completely halted \citep{hunter1998}. These galaxies are noted for their rather chaotic spatial distributions of their star forming regions, typically being asymmetrical and clumped on large scales \citep[e.g.][]{hodge1975}. The extremely low surface brightnesses of many of these systems makes detection of the continuum stellar emission difficult, and SFDGs would likely appear as clumps of bright emission-line sources in H$\alpha$ surveys.

Additionally, types of dwarf galaxies are being found with very low masses (ultrafaint dwarfs, UFDs; \citealt{simon2019}) and very low surface brightness (ultradiffuse dwarfs, UDGs; \citealt{sandage1984, vandokkum2015}). Both types of galaxies represent the extreme faint end of the galaxy luminosity function, typically having luminosities fainter than $M_V = -7.7$ \citep{simon2019}. These dwarf galaxies tend to have ancient ages often consistent with star formation ending by reionization at $z \sim 6$ \citep{brown2014}. If there are star-forming analogues to these galaxy types in the local universe, they may be detectable through very deep and wide-field narrowband imaging \citep{vanderhulst1993,mcgaugh1994,schombert2011,cannon2011,cannon2018}. 

In an effort to explore these areas of low-density star formation, we have used Case Western Reserve University's Burrell Schmidt 24/36-inch telescope to perform the first deep, wide-field, multiline, narrowband observations of the nearby spiral galaxy M101 (NGC~5457, $D = \SI{6.9}{\mega\parsec}$; see \citealt{matheson2012} and references therein) and its group environment. We image in three different emission lines that are characteristic of star formation, H$\alpha$, H$\beta$, and [\ion{O}{3}], which allows us to more cleanly reject contaminants without the need for expensive follow-up spectroscopy. This strategy, combined with our large survey area ($\sim$\SI{6}{\square\deg}) and our deep photometric limit of \SI{5.7e-17}{\erg\per\second\per\square\centi\metre} (reaching SFR limits of \SI{1.7e-6}{\solarmass\per\year}) gives us a good census of extended and outlying star formation and faint star-forming dwarf galaxies over large areas in nearby groups. 

M101 was chosen for this survey because its nearby distance enables its properties to be studied in detail \citep{mihos2012,mihos2013,mihos2018,watkins2017}. M101 is also currently interacting with its satellite population, likely its massive satellite NGC~5474 as evidenced by its asymmetric disk \citep{beale1969,rownd1994,waller1997}. Given that interacting systems often display extended or outlying star-forming regions, the M101 Group could be an excellent case study to explore the conditions under which extended intragroup star formation is triggered. Additionally, M101 has been found to harbor ultradiffuse galaxies \citep{merritt2014,merritt2016,karachentsev2015,danieli2017,carlsten2019} and constraining the star-forming properties of these objects will aid in understanding star formation in low-density environments.

\section{Narrowband Imaging}

The narrowband imaging used here was taken over the course of three seasons using Case Western Reserve University's 24/36-inch Burrell Schmidt telescope, located at Kitt Peak in Arizona. Our narrowband imaging and data reduction techniques are described in detail in \citet{watkins2017} and summarized briefly here. 

The Burrell Schmidt images a $\ang{1.65}\times\ang{1.65}$ field of view onto a single 4096$\times$4096 back-illuminated CCD, yielding a pixel scale of \ang{;;1.45} \si{\per\pixel}. For each emission line studied (\ha, \hb, and \oiii), we image in two narrowband filters (see Table~\ref{narrowbandobs}) --- one centered on the emission line and another shifted $\approx$\SI{150}{\angstrom} off the emission line for continuum subtraction. Given the width of our filters (necessitated by the fast $f/3.5$ beam of the Schmidt; \citealt{nassau1945}), our \ha-on filter covers both \ha\ and the adjoining \nii$\lambda\lambda$6648,6583 lines, while the \oiii-on filter covers both lines of the \oiii$\lambda\lambda$4959,5007 doublet. In each filter, we image M101 using 55--71 images of \SI{1200}{\second} exposure time each, with each pointing dithered randomly by up to \ang{0.5}. All data was taken under dark, photometric conditions, with the \ha\ imaging taken in Spring 2014 (and described in detail in \citealt{watkins2017}), the \hb\ imaging in Spring 2018, and the \oiii\ imaging in Spring 2019. Flat fielding was done using a combination of twilight flats and offset night sky flats \citep[see][]{watkins2017}, and deep \SI{1200}{\second} observations of Regulus and Arcturus were used to model and correct for scattered light from bright stars in the field \citep[see][]{slater2009}. We subtract sky from each image using a simple plane fit to the background sky levels across the image (typically $\sim$\SI{100}{\adu}). After correcting for sky and scattered light, as well as for fringing from \ch{OH} sky lines in the H$\alpha$ on-band filter, each set of dithered images was then median-combined to yield a final on- and off-band master image of the field, representing a total exposure time of 18--24 hours in each filter.

We flux calibrate the narrowband imaging using three methods. First, we calibrate using observations of spectrophotometric standard stars \citep{massey1988} taken throughout the course of each night, solving for extinction coefficients and nightly zeropoints that are applied to each image. Second, we self-calibrate each image using the 100--150 stars in the field around M101 that have well-measured photometry from Sloan Digital Sky Survey (SDSS) imaging, applying a color-dependent offset between SDSS broadband filters and our narrowband filters synthesized using the \citet{pickles} Stellar Spectral Flux Library. Third, we self-calibrate each image using the the $\sim$100 SDSS spectroscopic point sources in the field, using the SDSS spectroscopy to synthesize calibrated AB magnitudes in our narrowband filters for zeropointing each image. These three independent techniques yielded zeropoints which agreed with one another to within $\pm$\SI{5}{\percent}, which we take as the uncertainty in our absolute photometric calibration.

\begin{deluxetable}{cccccc}
\tabletypesize{\scriptsize}
\tablewidth{0pt}
\tablecaption{Narrowband Imaging Datasets}
\tablehead{\colhead{Filter} & \colhead{$\lambda_0$} & \colhead{$\Delta\lambda$} &
\colhead{Exposure Time} & \colhead{ZP (flux)} & \colhead{ZP (AB)}}
\startdata
\ha-on & 6590 \AA & 101 \AA & $71 \times \SI{1200}{\second}$ & \num{5.61e-18} & 26.63\\
\ha-off & 6726 \AA & 104 \AA & $71 \times \SI{1200}{\second}$ & \num{5.50e-18} & 26.64\\
\hb-on & 4875 \AA & 82 \AA & $59 \times \SI{1200}{\second}$ & \num{7.65e-18} & 26.73\\
\hb-off & 4757 \AA & 81 \AA & $55 \times \SI{1200}{\second}$ & \num{7.91e-18} & 26.74\\
\oiii-on & 5008 \AA& 102 \AA & $67 \times \SI{1200}{\second}$ & \num{7.58e-18} & 26.91\\
\oiii-off & 5114 \AA & 101 \AA & $66 \times \SI{1200}{\second}$ & \num{7.37e-18} & 26.89\\
\enddata
\tablecomments{ZP (flux) converts 1 ADU to $\rm erg\ s^{-1}\ cm^{-2}$ in the master
images, while ZP (AB) converts to AB magnitudes.}
\label{narrowbandobs}
\end{deluxetable}

\section{Methods}

\subsection{Source Detection}

Our final imaging dataset from the Burrell Schmidt consists of the narrowband imaging described in Section~2, along with deep broadband imaging in Washington $M$, similar to Johnson $V$, and a modified (bluer) Johnson $B$ filter from \cite{mihos2013}. We start by detecting sources on the H$\alpha$ on-band image, using \texttt{astropy}'s \texttt{PhotUtil} package, specifically the \texttt{segmentation} module \citep{bradley2019}. This program detects sources as objects that have a minimum number of connected pixels that are each greater than the background threshold value. In our case, the threshold value above which pixels would be marked as a detection was $3\sigma$ above the background level on the H$\alpha$ on-band image after a two pixel Gaussian smoothing. This sigma-clipping is conceptually similar to \texttt{SExtractor}'s $\kappa\sigma$ clipping \citep{sextractor}. 

In order to avoid detecting random noise spikes or star-forming objects well inside known bright galaxies, we masked several regions in the H$\alpha$ on-band image. First, we masked a \SI{750}{\pixel} ($\sim$\ang{;19;}) border around the image where the background noise becomes dominant. We also masked stars in the Tycho-2 Catalog \citep{hog2000} brighter than $B_T = 12.5$. Circular masks were applied to many of the galaxies in our survey area corresponding to twice the $R_{25}$ isophotal radius, taken from the RC3 catalog \citep{rc3}. This mask size follows from what has been done previously by other authors in defining the boundary beyond which lies outlying \ion{H}{2} regions \citep{werk2010}. The only exception to this was M101 itself; \cite{mihos2013} showed that the $R_{25}$ reported in the RC3 significantly overestimates M101's $\mu_B = 25$ isophotal radius by as much as a factor of two. We utilized the areal-weighted $R_{25} = \ang{;8;}$ for M101 reported in \cite{mihos2013}. 

Having masked these regions, we then create a two-dimensional background object to calculate the background sky level and its uncertainty. The sky level was estimated in boxes of $100 \times 100$ pixels with filter sizes of $10 \times 10$ pixels. Then, as mentioned before, we detected sources that were $3\sigma$ above the background, resulting in \num{32439} sources. We used the default parameters of 32 multi-thresholding levels and a contrast of $0.001$ to deblend close or overlapping sources. The \texttt{segmentation} module uses a combination of multi-thresholding and watershed segmentation to separate overlapping sources, which, given the parameters above, results in a segmentation map with a total of \num{35308} sources. For context, \texttt{SExtractor} utilizes only a multi-thresholding technique to deblend sources \citep{sextractor}. 

Using this segmentation map to define each source, we calculated photometric and structural quantities for each source in each of the narrowband images as well as in the broadband imaging of \cite{mihos2013}. Many of these were default calculations for \texttt{segmentation}, including positions and fluxes. We also calculated photometric errors, signal-to-noise, and AB magnitudes for each object in each filter. In what follows, the magnitude in each filter will be written as $m_{\lambda_0}$, where $\lambda_0$ is the central wavelength of the filter. Additionally, we calculate flux differences, $\Delta f = f_{\text{on}} - f_{\text{off}}$, and emission line equivalent widths (EWs) in each pair of filters. We again note that our H$\alpha$ on-band filter bandpass includes the [\ion{N}{2}]$\lambda$,$\lambda$6549,6583 doublet, thus measuring $\text{H}\alpha + [\text{\ion{N}{2}}]$. We have corrected these fluxes by adopting the emission line ratio of [\ion{N}{2}]/H$\alpha = 0.33$ \citep{kennicutt1992,jansen2000}.

Given that the net flux in a filter pair can be either positive or negative, we express the net flux in each band using asinh magnitudes \citep{lupton1999}: 
\begin{equation*}
    m(\Delta f) = \text{ZP} - 2.5\log(b) - a\arcsinh\left(\frac{\Delta f}{2b}\right)
\end{equation*}
Here, $a \equiv 2.5\log e = 1.08574$ is Pogson's ratio \citep{pogson1856}, $\Delta f$ is the flux difference of the source in a particular filter, and the zero point, ZP, was chosen so that objects with $\Delta f = 0$ had zero magnitude. This system has the benefit that it can be calculated for any flux value, negative or positive, and behaves smoothly as the flux drops through zero. The softening parameter, $b$, was chosen to be the flux of an object with $\text{S/N} = 1$. The softening parameter and zero point were calculated per filter set; specific values can be seen in Table \ref{b_zp}. In this magnitude system, objects with positive net flux (``in emission'') will have numerically negative magnitudes, while objects with negative net flux (``in absorption'') will have positive magnitudes. Going forward, we will refer to these ``net flux'' magnitudes as $m(\Delta f)$ in each spectral line.

\begin{deluxetable}{lcc}
\tabletypesize{\scriptsize}
\tablewidth{0pt}
\tablecolumns{3}
\tablecaption{Asinh Magnitude Parameters\label{b_zp}}
\tablehead{\colhead{Filter Set} \vspace{-0.2cm} & \colhead{$b$} & \colhead{ZP}  \\ 
& \colhead{{[\SI{e-17}{\erg\per\second\per\centi\metre\squared}]}} & }
\startdata
H$\alpha$ & \num{3.8} & -\num{41.05} \\
H$\beta$ & \num{5.4} & -\num{40.67} \\
{[\ion{O}{3}]} & \num{5.3} & -\num{40.69}
\enddata
\end{deluxetable}

We also cross-matched our source list with those objects in the SDSS DR16 \citep{ahumada2020} and with \emph{Gaia} Early Data Release 3 (EDR3; \citealt{gaiaedr3}). The \emph{Gaia} data will aid in rejecting interlopers in our dataset, while the SDSS data provides additional photometry and structural information for our final sample of objects. 

At this point, of those \num{35308} sources detected, \num{5442} ($\sim$\SI{15}{\percent}) were detected ``in emission'' (i.e.\ have positive net flux at the $3\sigma$ level) in H$\alpha$ and form the starting sample for our emission line source catalog. Only \num{450} ($\sim$\SI{1}{\percent}) were detected ``in emission'' in all of the three narrowband filters. However, as we will show next, many of these detections were not true narrowband emission features. 

\subsection{Investigating Potential Sources}

At this point in our detection routine, we have a source catalog of any astronomical source that has excess emission in H$\alpha$. This could include objects such as \ion{H}{2} regions or dwarf galaxies in the nearby universe, background emission line objects, or objects such as M stars with molecular absorption bands in our filters. 

\begin{figure*}
\epsscale{1.2}
\plotone{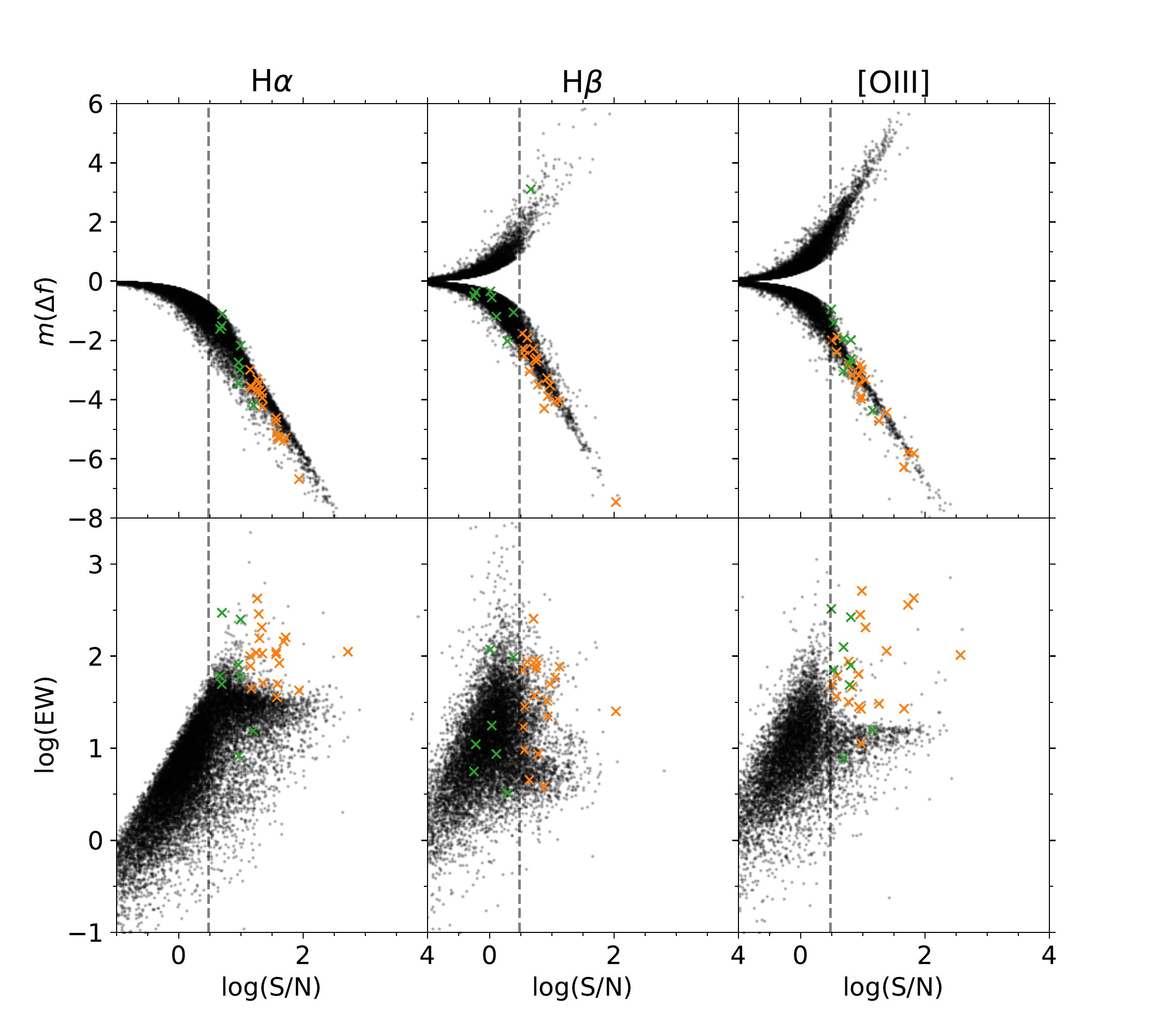}
\caption{Top row: The H$\alpha$, H$\beta$, and [\ion{O}{3}] asinh net flux magnitude as a function of signal-to-noise for each source observed in emission in H$\alpha$ (negative $m(\Delta f)$). Bottom row: The narrowband equivalent width (EW) as a function of signal-to-noise in each filter set. In both rows, the vertical dashed line indicates $\text{S}/\text{N} = 3$; sources at lower S/N are considered undetected in our analysis. Black points show all sources, orange crosses are sources detected in emission in all three lines, and green crosses are sources detected in two lines (H$\alpha$ and [\ion{O}{3}]). See text for details. \label{asinh}}
\end{figure*}

This is illustrated in Figure~\ref{asinh} where we plot our ``net flux'' magnitudes and EWs against the signal-to-noise of each source in each filter set. Since we are attempting to detect \ion{H}{2} regions, we select for only those sources with net positive H$\alpha$ flux differences, i.e.\ negative ``net flux'' magnitudes. Most of our sources have very low signal-to-noise, so we use only those sources detected with $\text{S/N} > 3$. These plots are conceptually similar to those in Figure 2 of \cite{kellar2012} used to search for compact emission line sources, except that in place of their ``ratio'' quantity, we use a true signal-to-noise, and in place of their magnitude difference, we use our ``net flux'' magnitudes. 

The most notable aspect of Figure~\ref{asinh} is that both the H$\beta$ and [\ion{O}{3}] filters have sources in absorption despite selecting for objects that are only in emission in H$\alpha$. We investigate this further using spectral synthesis of various objects -- stars, nearby galaxies, and high redshift objects -- through our filters to assess their behavior in our sample selection criteria. 

\begin{figure}
\epsscale{1.2}
\plotone{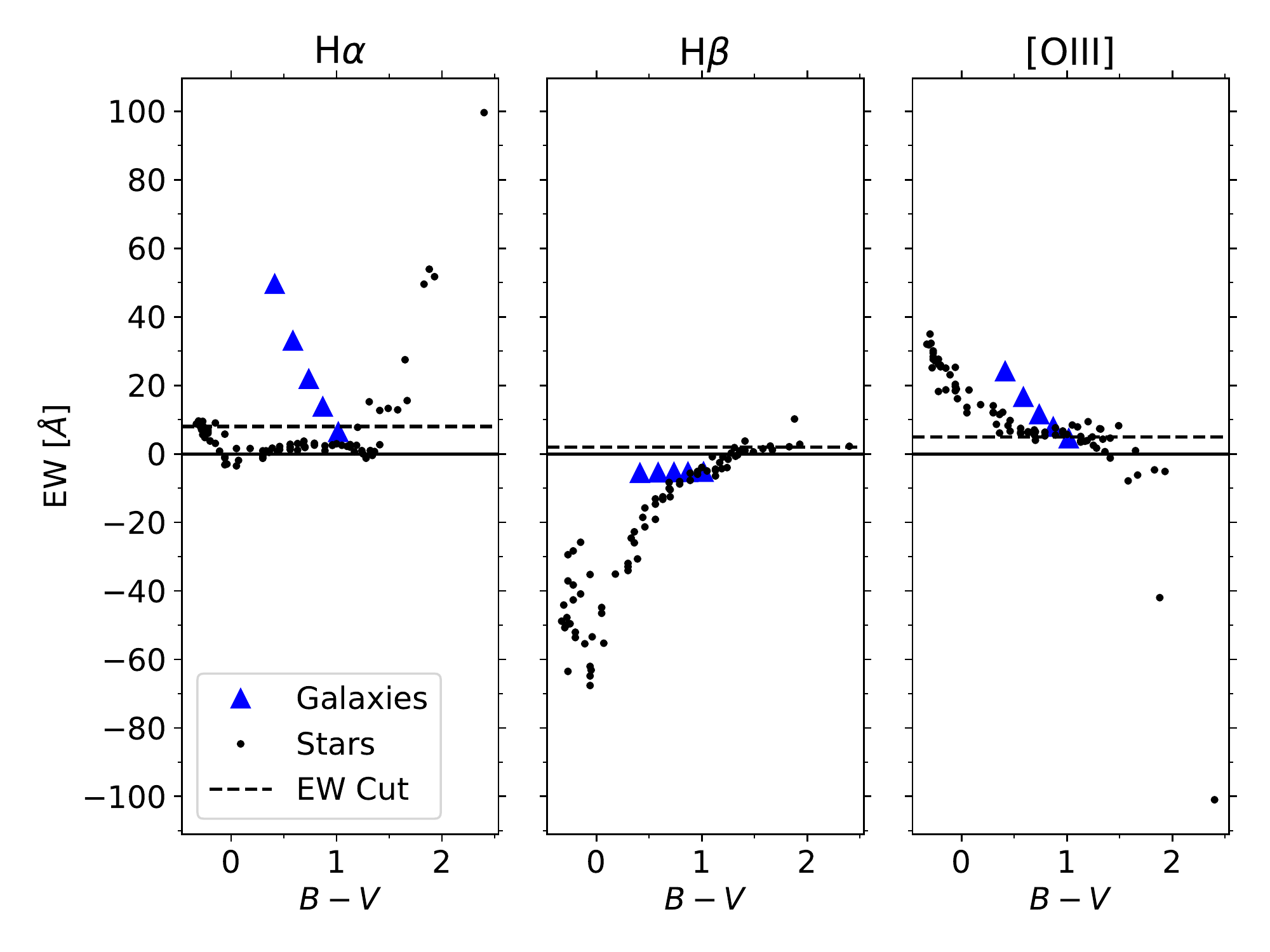}
\caption{Synthesized equivalent width as a function of color for stars in the \protect\cite{pickles} Stellar Spectral Flux Library (black points) and $z=0$ SDSS galaxy spectral templates (blue triangles). The horizontal dashed line indicates the equivalent width selection cut for our emission line sample in each filter pair. \label{stars_gals}}
\end{figure}

To test how other stars would behave in our filters, we synthesized spectra from the Stellar Spectral Flux Library \citep{pickles} through our filters. The Stellar Spectral Flux Library was chosen because it offers a wide distribution of stars of various spectral types, luminosity classes, and metallicity while also covering a large wavelength range with uniform $R \sim 500$ spectral resolution. For nearby galaxies, we adopt the SDSS DR5 spectral template for different galaxy types, synthesizing their EWs at zero redshift, such that the emission lines fall within our filters. The synthesized EWs in each filter are shown as a function of $B-V$ color in Figure~\ref{stars_gals}. 

It is clear from Figure~\ref{stars_gals} that stars bluer than $B-V\sim 1$ will broadly mimic H$\alpha$ emission, H$\beta$ absorption, and [\ion{O}{3}] emission. Meanwhile, stars redder than $B-V\sim 1$ will broadly mimic H$\alpha$ emission, H$\beta$ emission, and [\ion{O}{3}] absorption. However, these are not true narrowband absorption or emission features, but rather are a result of how our narrowband filters sample features in the stellar continuum. Because continuum starlight produces these low level ``pseudo-emission'' features, in our search for true star-forming signatures, we ignore any object with EW values lower than $\text{EW}(\text{H}\alpha) = \SI{8}{\angstrom}$, $\text{EW}(\text{H}\beta) = \SI{2}{\angstrom}$, and $\text{EW}(\text{[\ion{O}{3}]}) = \SI{5}{\angstrom}$ as shown by the dotted lines in Figure~\ref{stars_gals}. Given the relative weaker strength of the H$\beta$ line compared to H$\alpha$, we make a separate distinction between sources with all three filters in emission, satisfying all the EW cuts above, and sources with H$\alpha$ and [\ion{O}{3}] filters in emission, satisfying only those two EW cuts above. These two groups will be called the three-line sample and two-line sample, respectively. Making those cuts reduces our source catalog from \num{35308} sources to \num{147} sources with $\text{S/N} > 3$ (95 in the three-line sample and 52 in the two-line sample). 

At zero redshift, galaxy SEDs show the expected trend between line emission and color: bluer late-type galaxies are stronger in EW than the redder early-type galaxies due to young stellar populations in the former. The EW cuts made above will only cut out the reddest galaxies with little to no H$\alpha$ or [\ion{O}{3}] emission. Therefore these cuts will not negatively impact our search for star-forming galaxies even at relatively low EW. 

Finally, we investigate contamination of our sample due to high redshift objects in the M101 field. In the H$\alpha$ survey of \cite{kellar2012}, \SI{37}{\percent} of their detected sources were higher redshift objects where the [\ion{O}{3}] line was redshifted into their H$\alpha$ filter. \cite{watkins2017} detected a handful of background galaxies and quasars in their H$\alpha$ sample, so the possibility of detecting high redshift objects in any one filter is strong. While our use of three narrowband filters significantly reduces the chance of background contaminants, there are still regions of redshift space where bluer emission lines can redshift into our filters. 

\begin{figure}
\epsscale{1.2}
\plotone{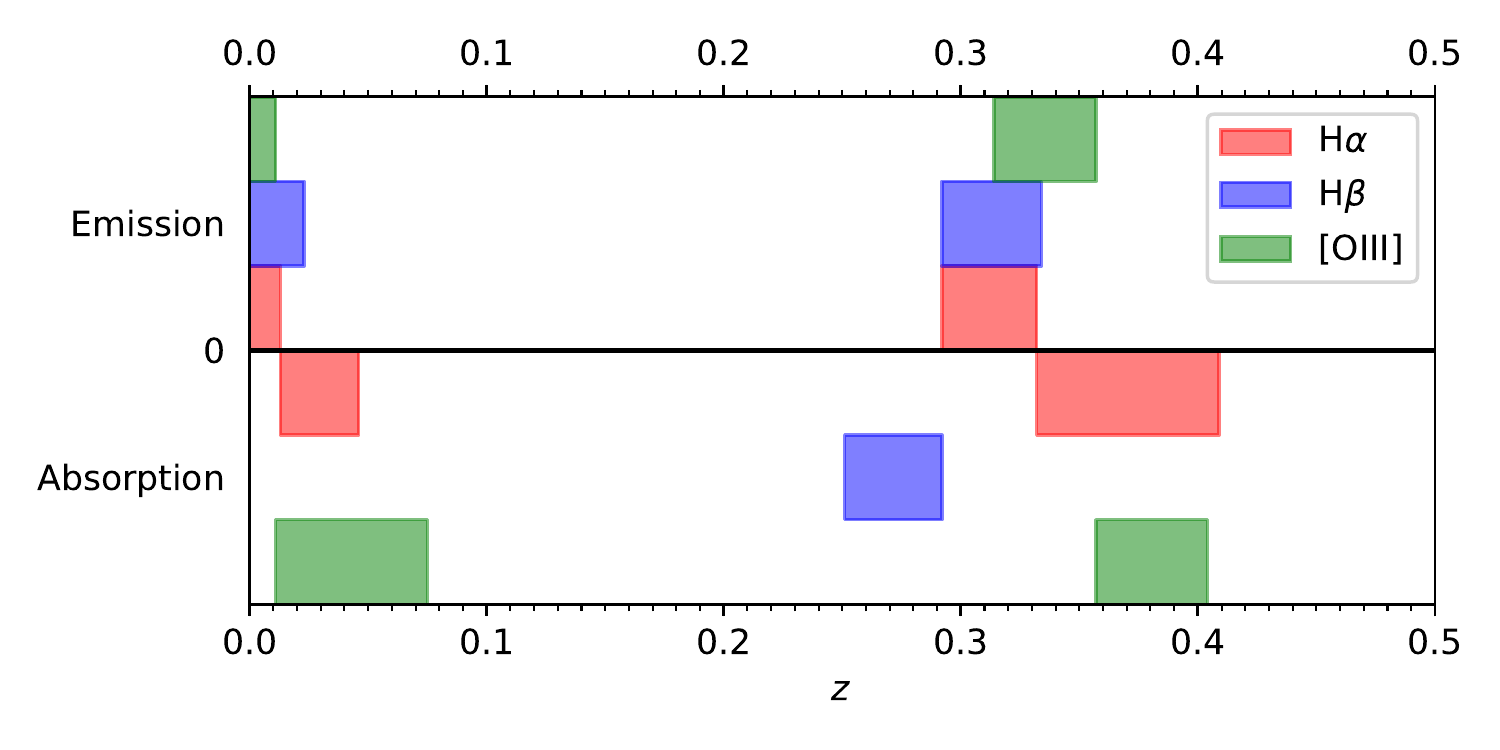}
\caption{Regions of redshift space where bright nebular emission lines manifest emission or absorption signatures as they redshift through our narrowband filters. Note that the legend indicates the narrowband filter pair, not the redshifted emission line. \label{redshift}}
\end{figure}

Figure~\ref{redshift} shows where common bright emission lines in star-forming galaxies can shift through our filters over the range of redshift $0 \leq z \leq 0.5$. At $z < 0.01$, all three filters will appear in emission. If we select for only those redshift ranges that satisfy the EW cuts above, i.e.\ objects detected in the three- or two-line samples, then emission line sources in the narrow redshift window $0.314 < z < 0.332$ can also potentially contaminate our samples. Here, the H$\alpha$ emission line has redshifted out of our filters entirely, while the [\ion{O}{3}]$\lambda\lambda4959{,}5007$ doublet has redshifted into the H$\alpha$ filters. Similarly, the H$\beta$ line has redshifted into the [\ion{O}{3}] filters and the [\ion{O}{2}]$\lambda\lambda3727{,}3729$ lines have redshifted into the H$\beta$ filters. Due to the small size of the Burrell Schmidt, star-forming galaxies at higher redshift are unlikely to be detected at all, but bright high redshift AGN may still produce some contamination of the sample. However, the rarity of such objects makes them unlikely to be present in large numbers in our sample. 

\subsection{Removing M Stars}

\begin{figure}
\epsscale{1.2}
\plotone{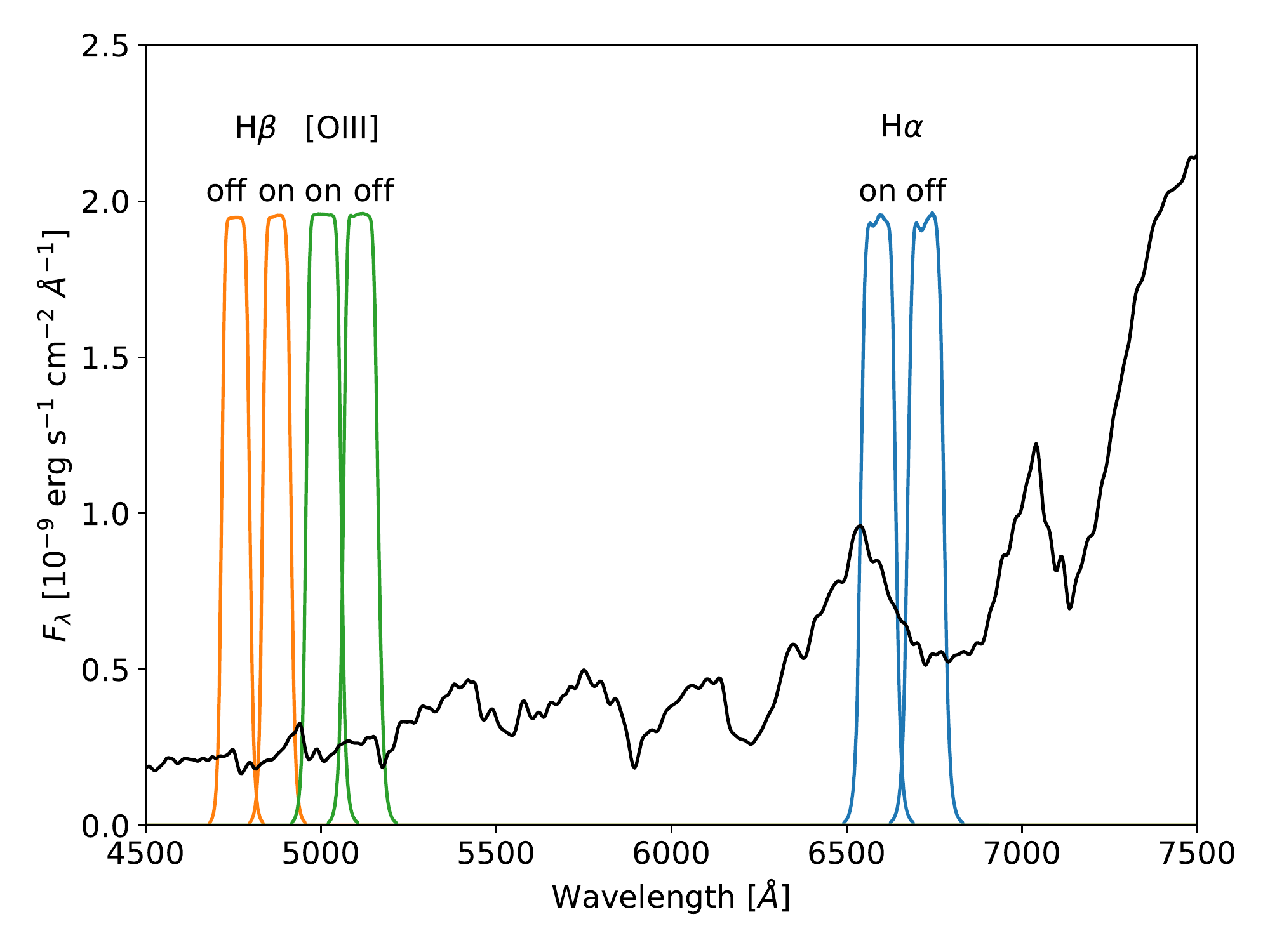}
\caption{The spectrum of a M5V star from \protect\cite{pickles} (solid black line) overplotted with our narrowband filter transmission curves (colored lines), showing how features in the stellar continuum of M stars produce pseudo-emission signatures in our narrowband filters.\label{Mstar}}
\end{figure}

While the equivalent width cuts significantly reduce contamination due to Milky Way stars, M stars continue to pose a particular challenge. The reason for this can be seen in Figure~\ref{Mstar}, which shows the spectrum of a M5V star from the Stellar Spectral Flux Library \citep{pickles}. Overplotted are the filter transmission curves of our narrowband filters. The H$\alpha$ off-band filter lies in a \ch{TiO} molecular absorption trough, producing net emission in the measured H$\alpha$ flux. Other features in the complex stellar continuum mimic emission in our H$\beta$ and [\ion{O}{3}] filters as well. The abundance of Galactic M stars at the faint end of the stellar mass function suggests many of our ``emission line'' detections will be M star contaminants needing to be removed.  

One possible method to removing M stars from our detections is to use some definition of compactness to distinguish between stellar point sources and extended \ion{H}{2} regions. Indeed, such a measure of compactness, the Gini index \citep[e.g.][]{lotz2004}, is part of the standard calculations made by the \texttt{segmentation} routine. However, given the FWHM of the Schmidt imaging (\ang{;;2.2}) and the assumed distance to M101, we would only be able to resolve objects larger than \SI{75}{\parsec}. The Str\"{o}mgren sphere radius of an \ion{H}{2} region powered by an O9 star is $\sim$\SI{30}{\parsec}, illustrating that some \ion{H}{2} regions would be unresolved in our imaging. Therefore, using compactness to distinguish between stars and \ion{H}{2} regions would bias us against the detection of smaller \ion{H}{2} regions in our survey.

\begin{figure}
\epsscale{1.2}
\plotone{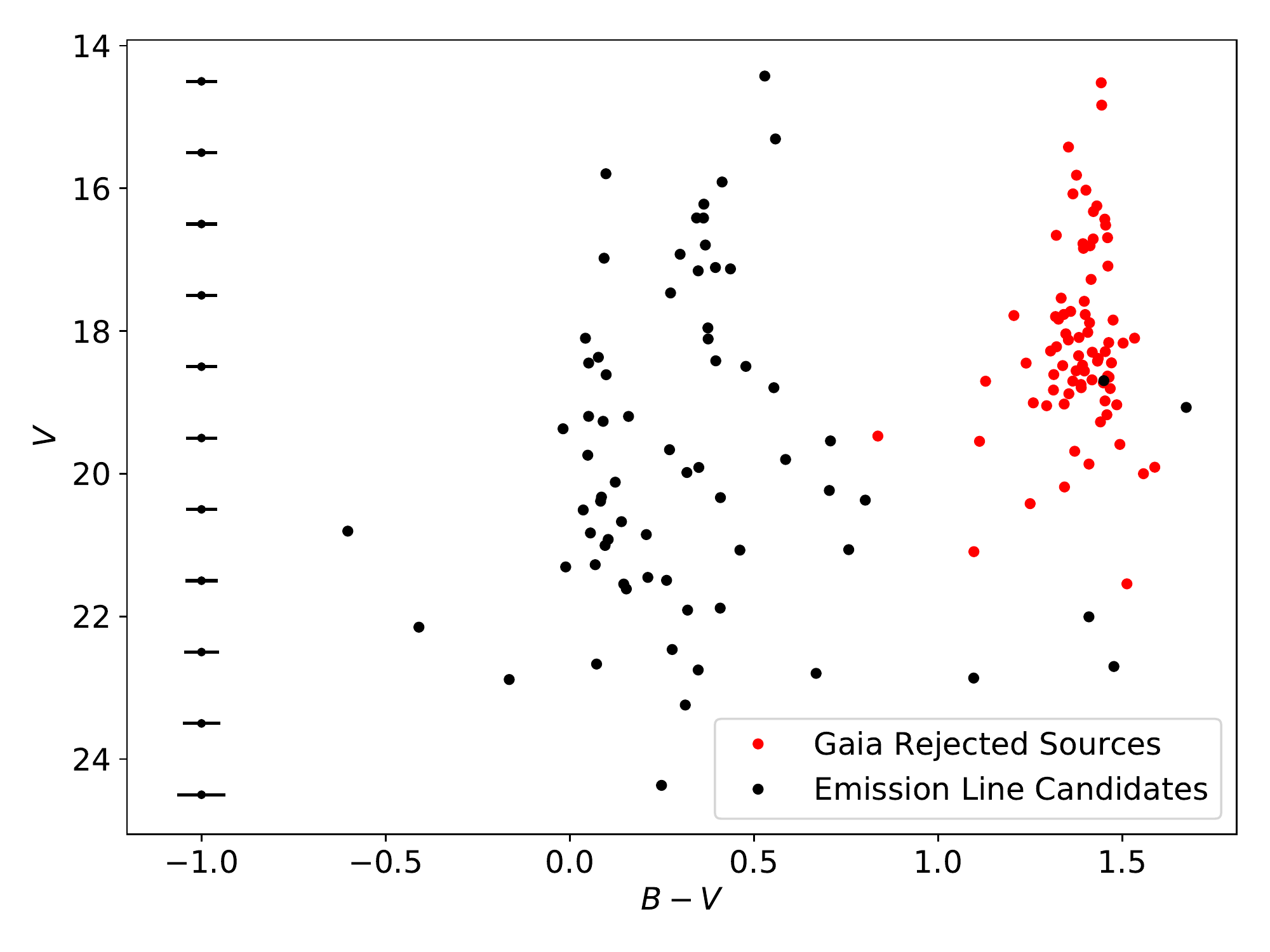}
\caption{Broadband color-magnitude diagram for sources in both the three- and two-line samples. The red points are those objects rejected by the \emph{Gaia} parallax and proper motion cuts. Characteristic photometric errorbars are shown, which include both uncertainties in the source photometry and in the overall photometric zeropoints. \label{bmv_gaia}}
\end{figure}

Instead, to remove any remaining Milky Way stars from our emission line sample, we cross-match our sources with the \emph{Gaia} EDR3 catalog. Since star-forming objects in the M101 Group should have no detectable parallax or proper motion, we cut any cross-matched source with a $3\sigma$ detection of parallax or proper motion. The results of these cuts are shown in Figure~\ref{bmv_gaia}. Nearly all of the sources with $B-V > 1$ are rejected by the \emph{Gaia} cuts on parallax and proper motion, indicating that they are Galactic M stars. There are a few red objects that do not appear in the \emph{Gaia} catalog at all. Of the brighter ones at $V\sim19$, the object at $B-V \sim 1.5$ is cataloged by SDSS as a galaxy with a photometric redshift of $z = \num{0.188 \pm 0.034}$, while redder one at $V\sim19$ and $B-V \sim 1.7$ is a point source in the SDSS imaging. Of the three fainter objects at $V\sim 22-23$, the bluest one at $B-V = 1.1$ is cataloged as a galaxy in the SDSS imaging, with a photometric redshift of $z=\num{0.388 \pm 0.137}$. Given the uncertainty in the photometric redshift, this object may be an example of a background contaminant leaking into our sample through the redshift window shown in Figure~\ref{redshift}. The other two faint red sources undetected by \emph{Gaia} appear as point sources in the SDSS imaging. Based on this combined analysis of \emph{Gaia} and SDSS properties, we reject all sources redder than $B-V=1$, including the five mentioned here that do not appear in the \emph{Gaia} catalog. 

\subsection{Additional Cuts}

Aside from contamination of the sample on the red end by Galactic M stars, Figure~\ref{bmv_gaia} shows a handful of very blue objects at $B-V < 0$. Cross-matching these sources against the SDSS imaging catalog reveals they are background sources in the M101 field. The two bluer objects are QSOs at redshift $z=1.34$ and $z=0.76$, where bright emission lines from [\ion{O}{2}] or \ch{Mg} have shifted into our filters. The reddest of the three is a background galaxy resolved in the SDSS imaging, but lacks any spectroscopic or photometric redshift. 

The high redshift objects mentioned above illustrate the need for a final cut to remove them. For objects in emission in the H$\beta$ filters, we utilize the $\text{H}\alpha/\text{H}\beta$ ratio as an additional interloper cut. For unobscured ionized gas, the Balmer decrement should be $\text{H}\alpha/\text{H}\beta = 2.86$ \citep{osterbrock1989} with higher ratios indicating higher extinction levels. Therefore objects in our sample which show much lower Balmer decrements are likely background sources where other emission lines have redshifted into the filters. Similarly, objects with anomalously high Balmer decrements are likely also contaminants, as star-forming galaxies in the local universe rarely get above a decrement of 8 \citep{dominguez2013}. We therefore keep only objects with $1 < \text{H}\alpha/\text{H}\beta < 8$ for the three-line sample. This cut removes four objects, including one of the QSOs noted above.

There is no similar calculation we can make for those sources in the two-line sample as they have no detected H$\beta$ emission. Additionally, depending on the specific redshift and combination of lines moving through our filters, it is possible that a background source could mimic a realistic Balmer decrement. To combat this, we cross-matched with SDSS, removing those sources for which SDSS has a spectroscopic or photometric redshift. This process removes 20 objects at higher redshift (8 QSOs and 12 galaxies), but also allows us to reject $z\sim0$ objects located just beyond M101. Two such examples of $z\sim0$ contaminants were detected: the blue compact dwarf galaxy SBS 1407+540 ($z = 0.0068$, $d = \SI{29.1}{\mega\parsec}$; \citealt{stepanian2005}) and the Magellanic-type irregular galaxy CGCG 272-015 ($z = 0.0071$, $d = \SI{30.4}{\mega\parsec}$; \citealt{ann2015}). After all of the cuts, we examined each object by eye to confirm the nature of the source; during this process one object was discarded due to it being an obvious blend, pairing a foreground M star with a background galaxy. 

Summarizing, we have split our source catalog into two groups: the three-line sample consists of those sources in H$\alpha$, H$\beta$, and [\ion{O}{3}], while the two-line sample is comprised of sources showing emission in H$\alpha$ and [\ion{O}{3}] only. Both samples have been cleaned of stellar contamination using a combination of EW cuts and \emph{Gaia} parallaxes or proper motions. Background contamination was removed by cross-matching with redshift estimates from SDSS imaging and spectroscopy. Finally, for the three-line sample, an additional cut was made on the observed Balmer decrement to reject non-physical values. Our final samples consist of 19 objects in the three-line sample and 8 objects in the two-line sample. 

\section{Analysis}

In this section, we investigate and describe the properties of the sources in the three-line and two-line samples. In total, the three-line sample contains 19 sources while the two-line sample contains 8 sources. Tables \ref{narrow_gold} and \ref{narrow_silver} list the narrowband properties of the sources in the three-line and two-line samples, respectively. The tables include a unique identifier for each source, the right ascension and declination (epoch J2000), whether that source is unresolved (U) or extended (E) in our images, the flux difference for each filter pair, the equivalent width (EW) for each filter pair, and the $\text{H}\alpha/\text{H}\beta$ and [\ion{O}{3}]/H$\alpha$ ratios. 

\begin{figure}
\epsscale{1.2}
\plotone{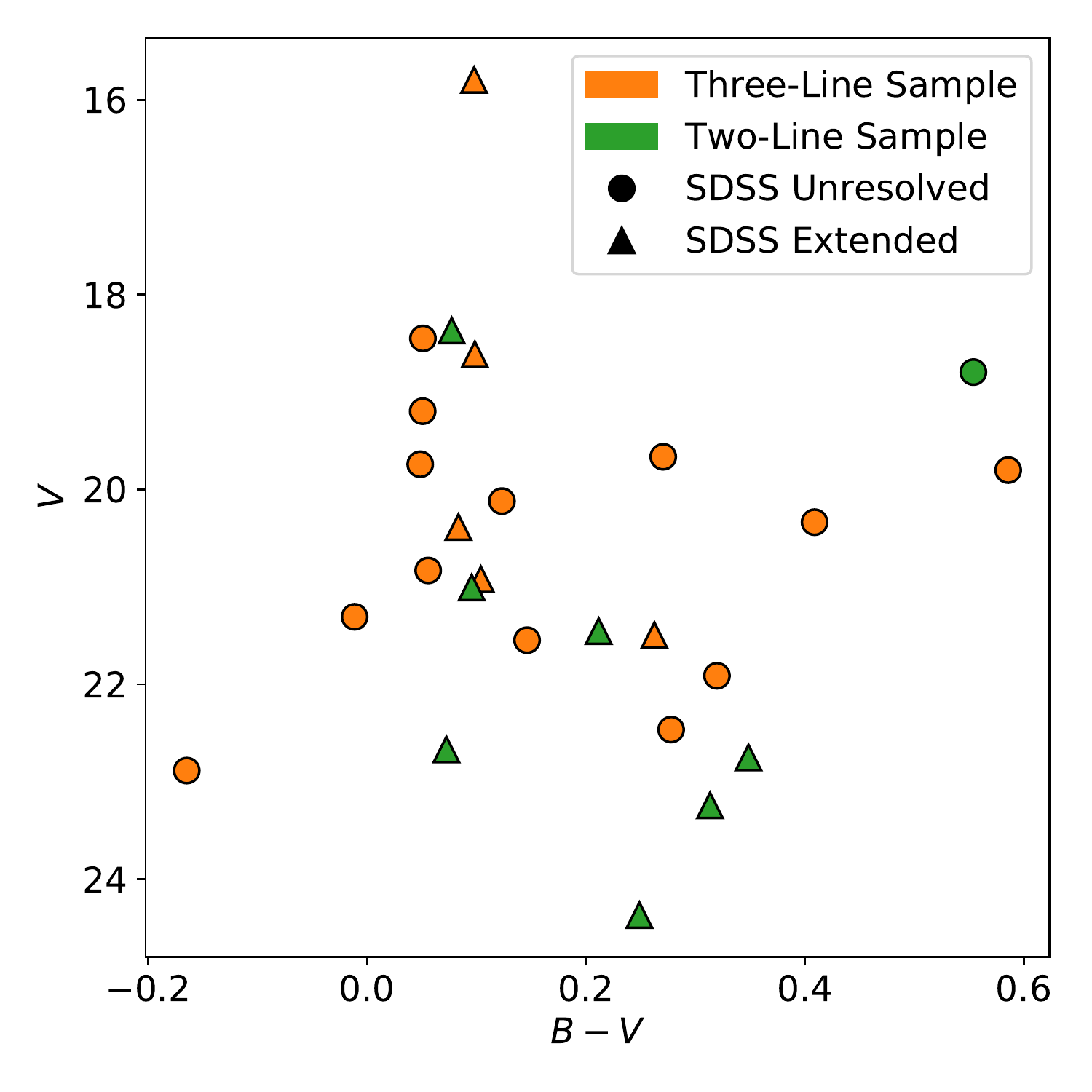}
\caption{Broadband color-magnitude diagram for sources in the final three-line (orange points) and two-line samples (green points). Circles show sources unresolved in SDSS imaging, while triangles show extended sources.\label{cmd}}
\end{figure}

Tables \ref{broad_gold} and \ref{broad_silver} list the broadband properties of the sources in the three-line and two-line samples, respectively. Included are the source identifier, the $V$-band magnitude, the $B-V$ color, the respective SDSS $ugriz$ photometry if available, and the \emph{GALEX} far ultraviolet (FUV) and near ultraviolet (NUV) magnitudes if available. The color-magnitude diagram of the sources is shown in Figure~\ref{cmd}. 

\begin{figure*}
\epsscale{1.2}
\gridline{\fig{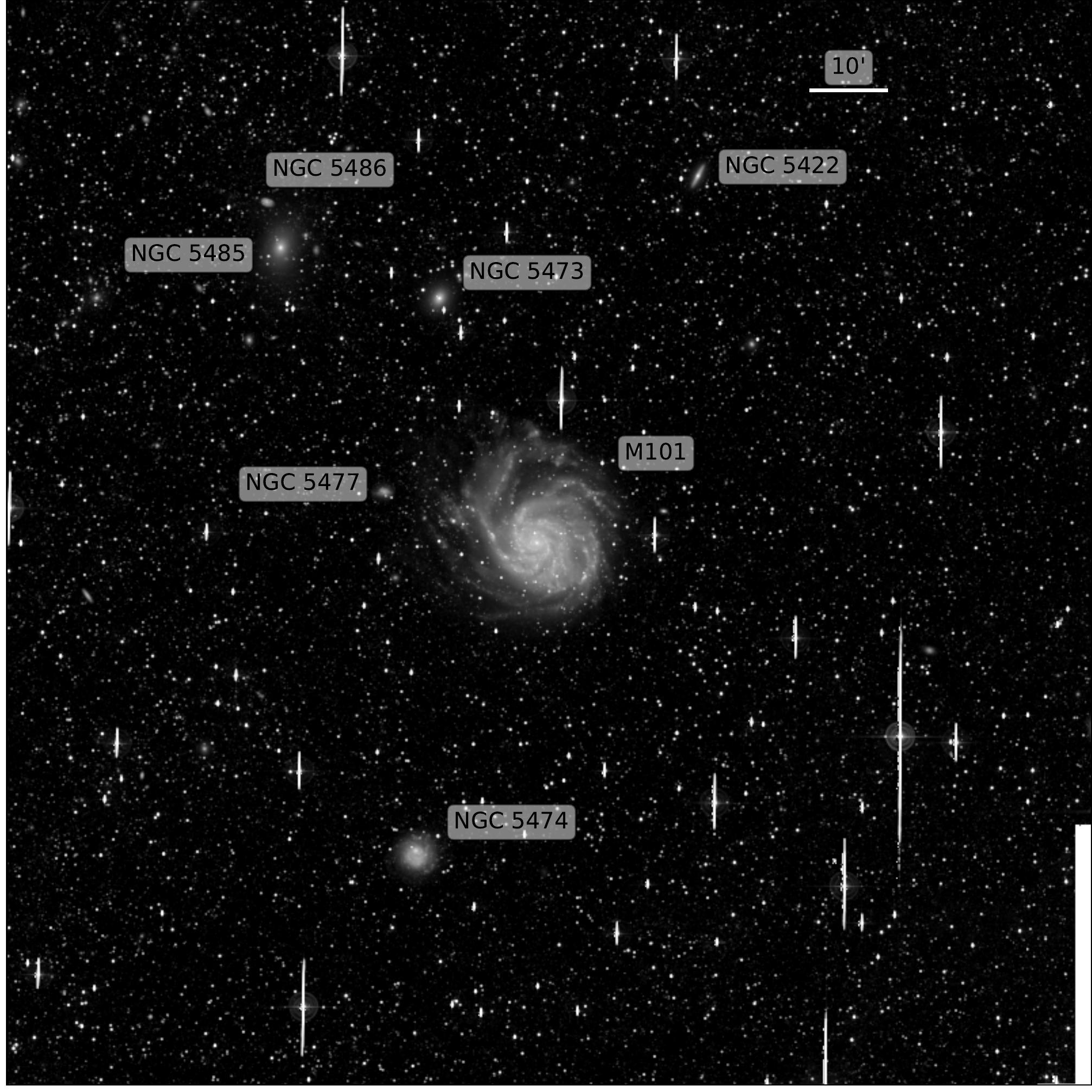}{0.5\textwidth}{(a)}
        \fig{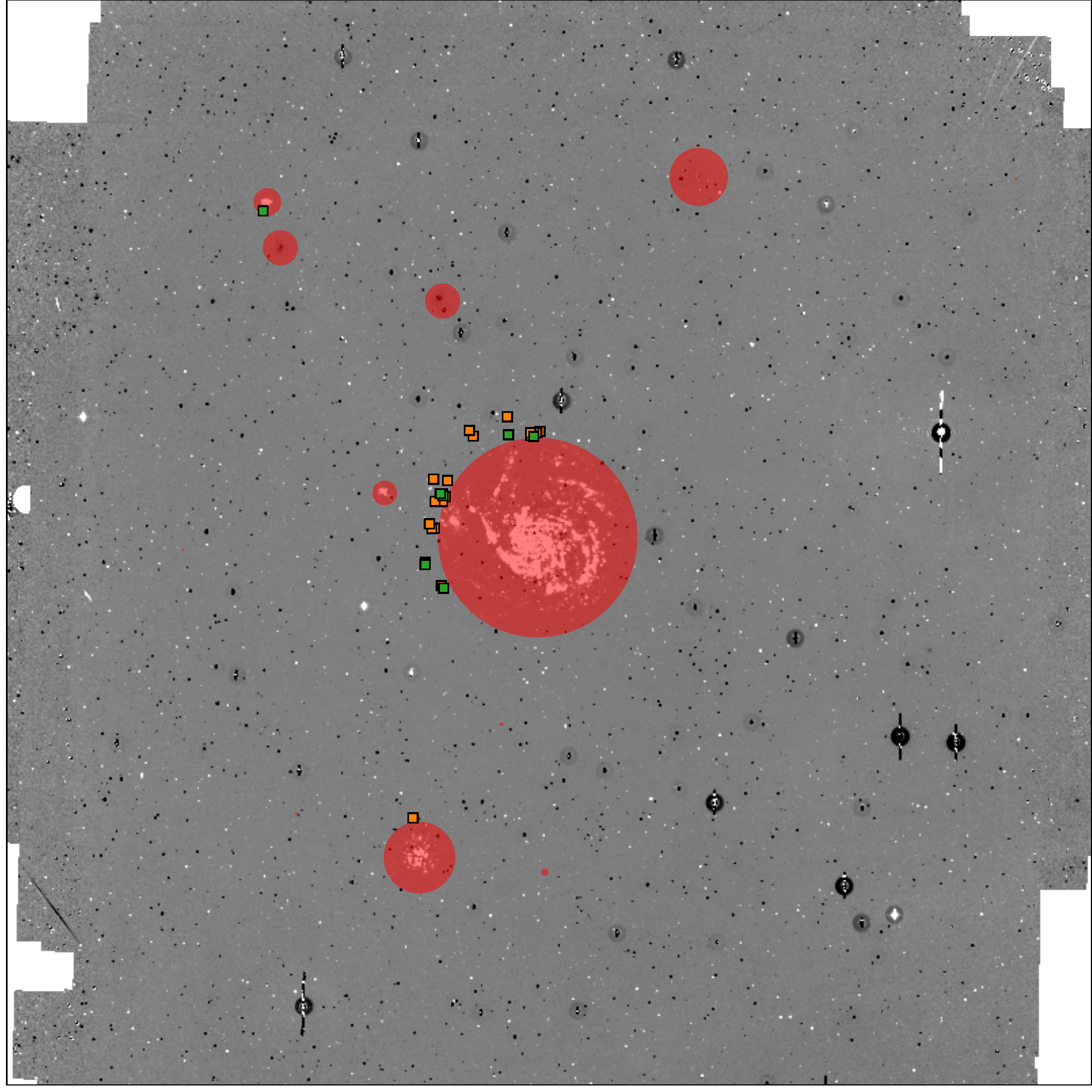}{0.5\textwidth}{(b)}}
\caption{(a) The broadband $B$ image with M101 and its companions, NGC~5474 and NGC~5477 labeled, as well as background galaxies also in the survey area. (b) The continuum-subtracted H$\alpha$ image. The red circles represent the masks used in our source detection algorithm. The orange and green squares represent the objects in our three-line and two-line samples, respectively. Both images measure $2.4 \times 2.4$ \si{\deg}. North is up and east is to the left.\label{images}}
\end{figure*}

\begin{figure*}
\epsscale{1.2}
\plotone{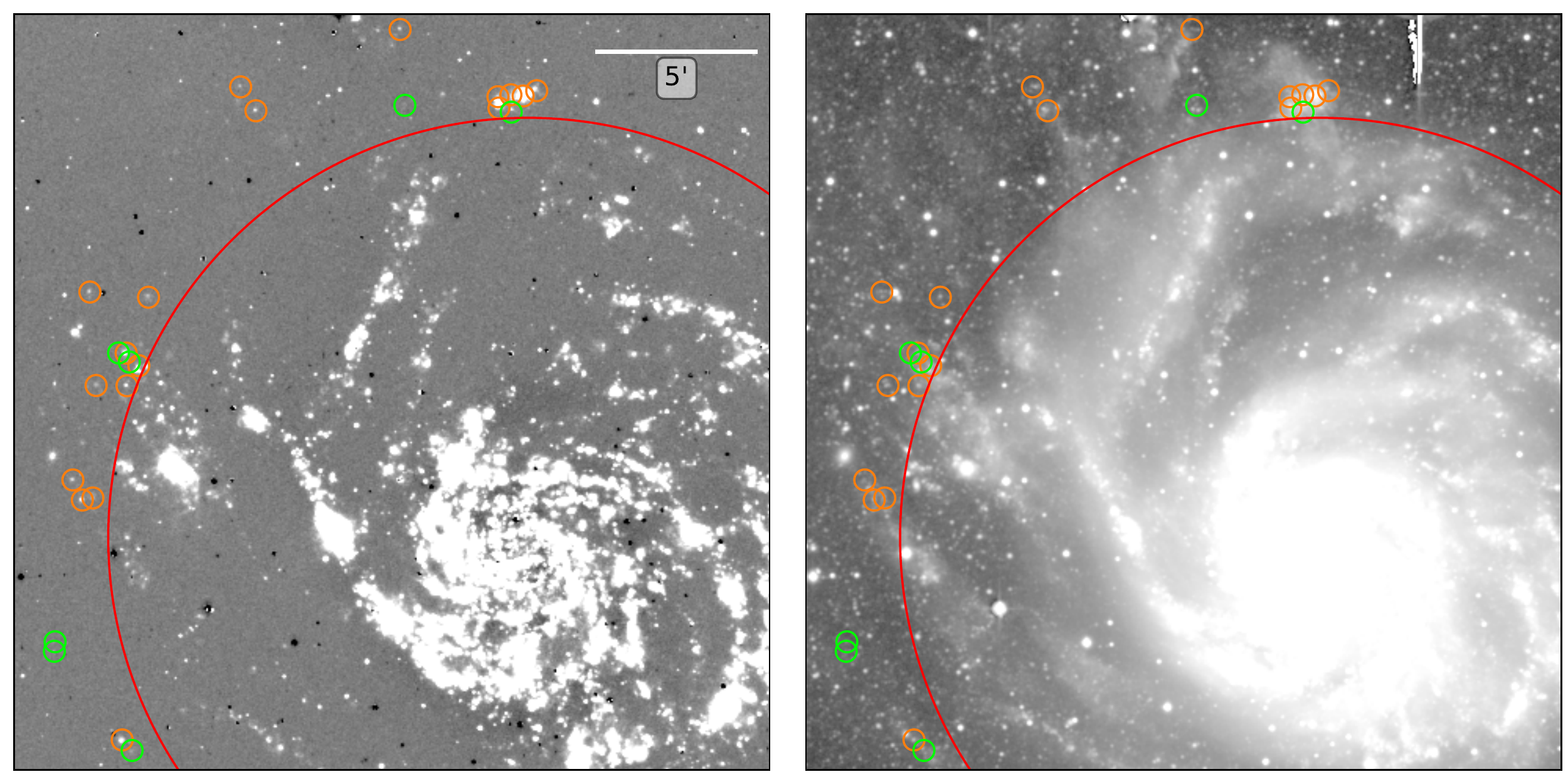}
\caption{A zoomed-in view of the detected regions surrounding M101. Left: the continuum-subtracted H$\alpha$ image. Right: the broadband $B$ image, stretched to show the faint structure of the outer disk. The orange and green circles with radii of \ang{;;20} are centered on the objects in our three-line and two-line samples, respectively. The red circle represents the size of the mask used in our source detection algorithm. Both images measure $24 \times 24$ \si{\arcmin}. North is up and east is to the left.\label{m101_detail}}
\end{figure*}

\begin{figure}
\epsscale{1.2}
\plotone{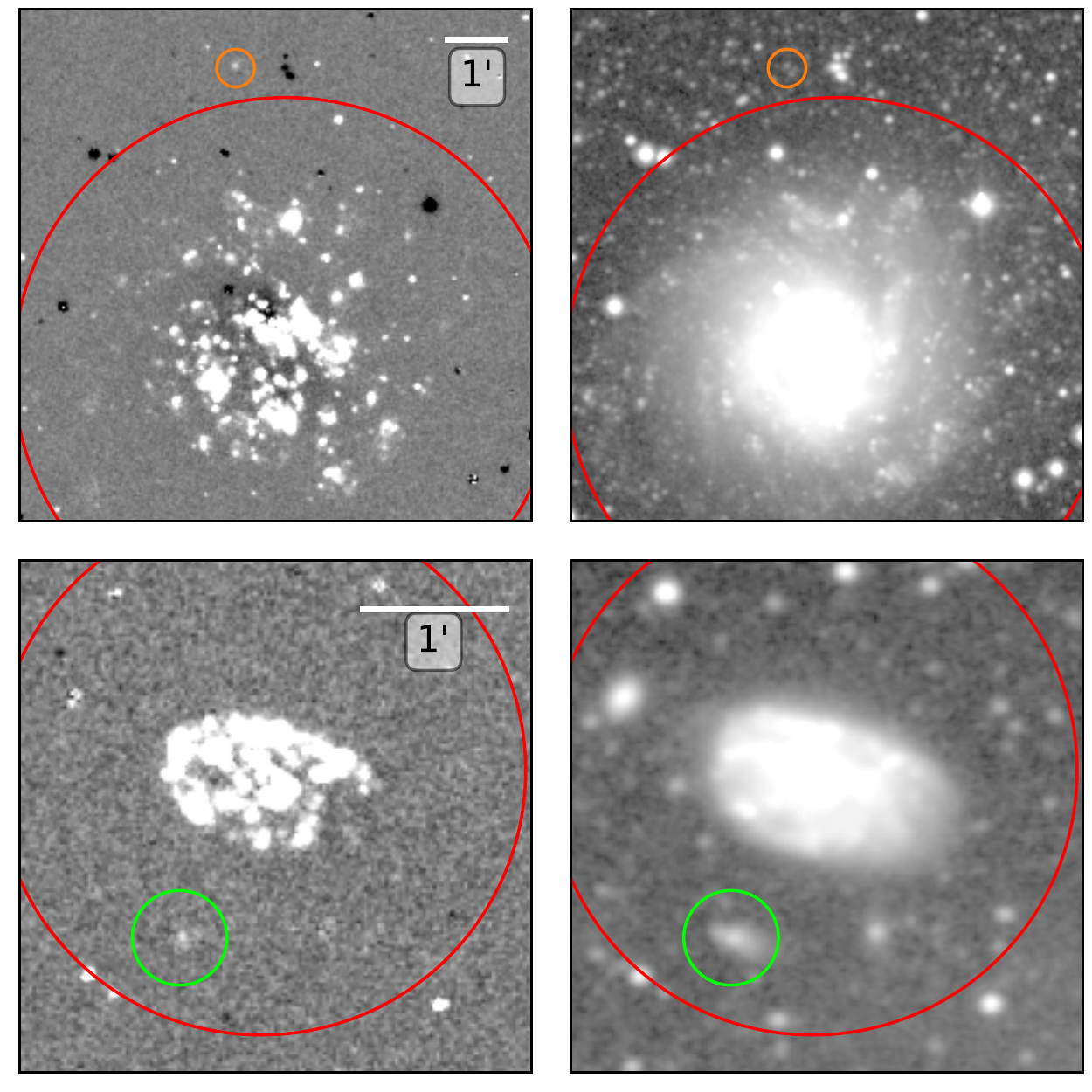}
\caption{A zoomed-in view of the detected regions near NGC~5474 (top row) and NGC~5486 (bottom row). Left: the continuum-subtracted H$\alpha$ image. Right: the broadband $B$ image, stretched to show the faint structure of the outer disks. Colored circles are the same as in Figure~\ref{m101_detail}. The images of NGC~5474 measure $9 \times 9$ \si{\arcmin} and the images of NGC~5486 measure $3.6 \times 3.6$ \si{\arcmin}. North is up and east is to the left. \label{dwarf_detail}}
\end{figure}

Briefly, we give a sense of how deep we have attained in fluxes and the range of equivalent widths investigated in each group. In the three-line sample, the faintest source has $V = 22.9$. The corresponding H$\alpha$ net flux is \SI{8e-16}{\erg\per\second\per\square\centi\metre}, which, at the \SI{6.9}{\mega\parsec} distance to M101 and using the SFR-$L(\text{H}\alpha)$ calibration of \cite{kennicutt2012}, would correspond to a SFR of \SI{2.4e-5}{\solarmass\per\year}. The three-line sample has a range of H$\alpha$ equivalent widths from \SIrange{35}{425}{\angstrom} with a median of \SI{110}{\angstrom}. In the two-line sample, the faintest source has $V = 24.4$. The corresponding H$\alpha$ net flux is \SI{9.3e-17}{\erg\per\second\per\square\centi\metre}, which at the distance of M101 corresponds to a SFR of \SI{2.8e-6}{\solarmass\per\year}. The two-line sample's H$\alpha$ equivalent widths range across \SIrange{8}{300}{\angstrom}, with a median of \SI{60}{\angstrom}. 

Figures~\ref{images}-\ref{dwarf_detail} show images of the detected sources and their surrounding environments. Source identifiers for each object were assigned based on their proximity to known objects in the survey area and their sample (see Figures \ref{images}-\ref{dwarf_detail}). For instance, sources labeled ``M101-3-\#'' are sources near M101 in the three-line sample. Additional sources were identified as belonging to NGC~5474 and NGC~5486. While the object associated with NGC~5486 lies just inside the galaxy's $2R_{25}$ isophotal radius, visual inspection shows the object to be a distinct source located well outside the galaxy's star-forming disk, so we have kept it in our sample. 

In the three-line sample, 18 (\SI{95}{\percent}) sources appear to be associated with the outer disk of M101 and 1 (\SI{5}{\percent}) appears to be associated with NGC~5474. In the two-line sample, 7 (\SI{88}{\percent}) sources appear to be associated with the outer disk of M101 and 1 (\SI{12}{\percent}) are associated with NGC~5486. 

\subsection{Structural Analysis}

Since we are looking for faint star-forming objects in the M101 Group, we begin by comparing the observed properties of our samples to the observed properties of known satellite galaxies in the Local Universe shifted to the M101 distance. We focus first on observed size, surface brightness, and apparent magnitude of our objects in the three- and two-line samples. 

Given the better resolution of SDSS, we assigned the sizes of our objects by utilizing the SDSS structural properties where able. For sources unresolved in SDSS imaging, we assigned a size equivalent to the FWHM of the SDSS $g$-band PSF (\ang{;;1.44}; \citealt{fukugita1996}). For resolved sources, we use the effective radius of the best-fit de Vaucouleurs or exponential profile as reported in the SDSS catalog. For objects not in SDSS, we measure the half-light radius directly from aperture photometry on the $V$-band Schmidt imaging. In this case, unresolved objects are again given an effective radius equivalent to the FWHM of the Schmidt imaging (\ang{;;2.2}) which corresponds to \SI{75}{\parsec} at M101. 

\begin{figure*}
\includegraphics[width=\textwidth]{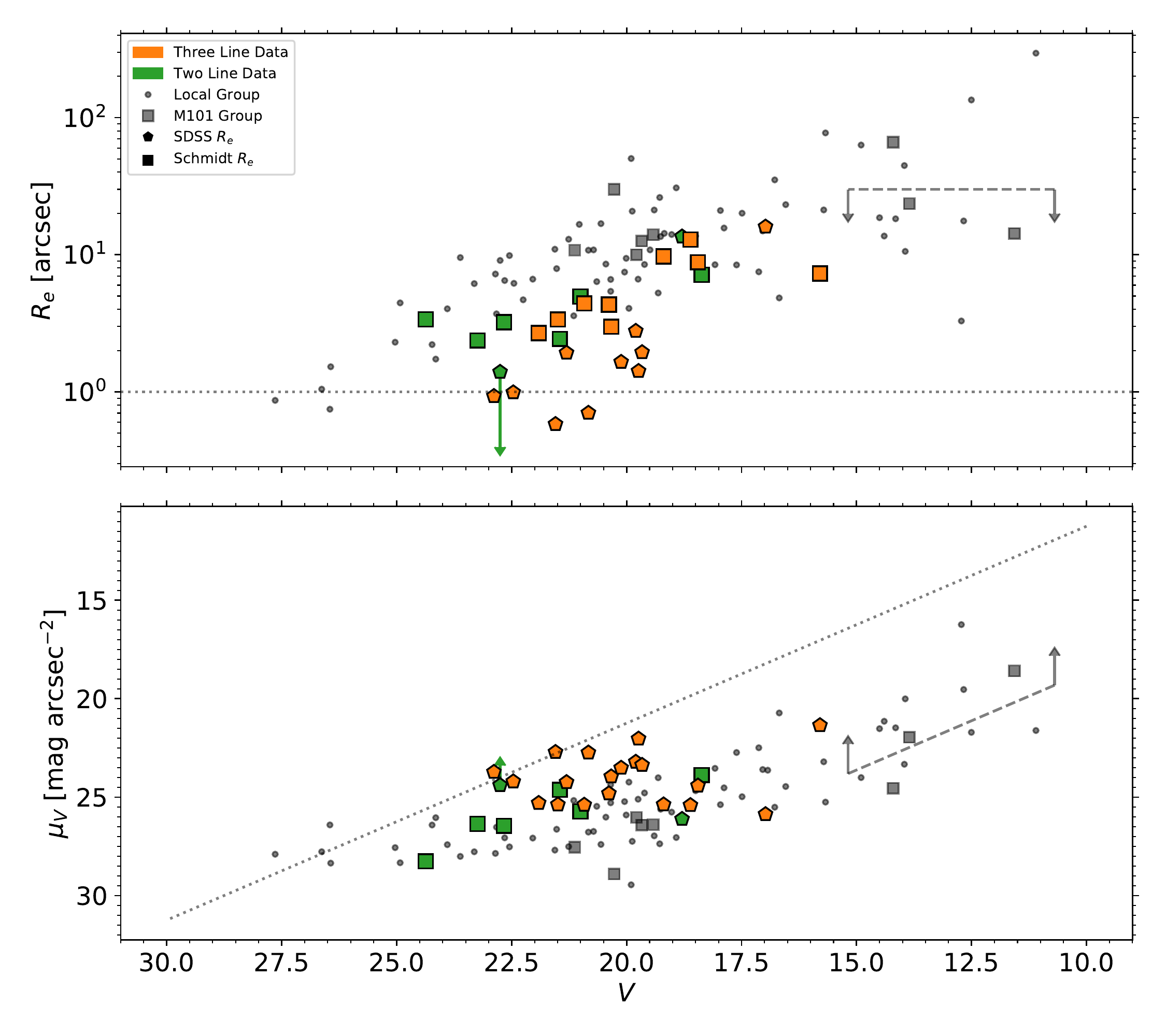}
\caption{Structural properties of our detected sources, compared to Local Group dwarfs shifted to the M101 distance. Top: Effective radius versus apparent magnitude. The dotted line indicates \ang{;;1} (\SI{33}{\parsec} at M101's distance). Bottom: Surface brightness versus apparent magnitude. The dotted line represents the limit on surface brightness for unresolved sources. Both the three- and two-line samples are plotted (orange and green markers, respectively). Pentagons denote objects whose properties were derived using SDSS data, while squares are sources with properties derived from our Burrell Schmidt imaging. Data for Local Group dwarfs are taken from \protect\cite{mcconnachie2012} and are shown as gray circles, while known M101 satellites are shown as gray squares \protect\citep{rc3,taylor2005,merritt2014,bennet2019,bellazzini2020}. The limits for \ion{H}{2} galaxies are shown as dashed gray lines \protect\citep{thuan1981}.\label{struc_all}}
\end{figure*}

We plot these quantities in the structural plots shown in Figure~\ref{struc_all}. The unresolved objects that have sizes equivalent to the SDSS or Schmidt PSFs are marked with arrows; these sizes only serve as upper limits to their true sizes. For context, the Local Group dwarf galaxies from \cite{mcconnachie2012} are also plotted. Effective radii for the LMC and SMC were taken from \cite{gallart2004} and \cite{massana2020}, respectively, while apparent $V$-band magnitudes were taken from \cite{rc3}. 

We also plotted several of the confirmed and candidate satellite galaxies of M101 according to \cite{carlsten2019}: NGC~5474, NGC~5477, Holmberg IV, DF1, DF2, DF3, DwA, and Dw9. There is no single paper that catalogs all of the photometric properties of the M101 satellite system; quantities for each galaxy were taken from a variety of sources \citep{rc3,taylor2005,merritt2014,bennet2019,bellazzini2020}. Finally, we also show the region where \ion{H}{2} galaxies would lie, using the definition of \cite{thuan1981}: $-18.0 < M_V < -13.5$ and sizes less than \SI{1}{\kilo\parsec}, which at M101 corresponds to an angular size of \ang{;;30}.

In general, the three- and two-line samples are separated into two groups: there is a group of larger objects that roughly follow the trends of known dwarf galaxies, and another set of smaller objects that do not. These latter objects are largely unresolved point sources, giving rise to an upper limit on size and lower limit on surface brightness as shown in Figure~\ref{struc_all}. Only one object lies near the region defined by \ion{H}{2} galaxies. We explore what this source might be by marking each object by associated galaxy.

\begin{figure*}
\includegraphics[width=\textwidth]{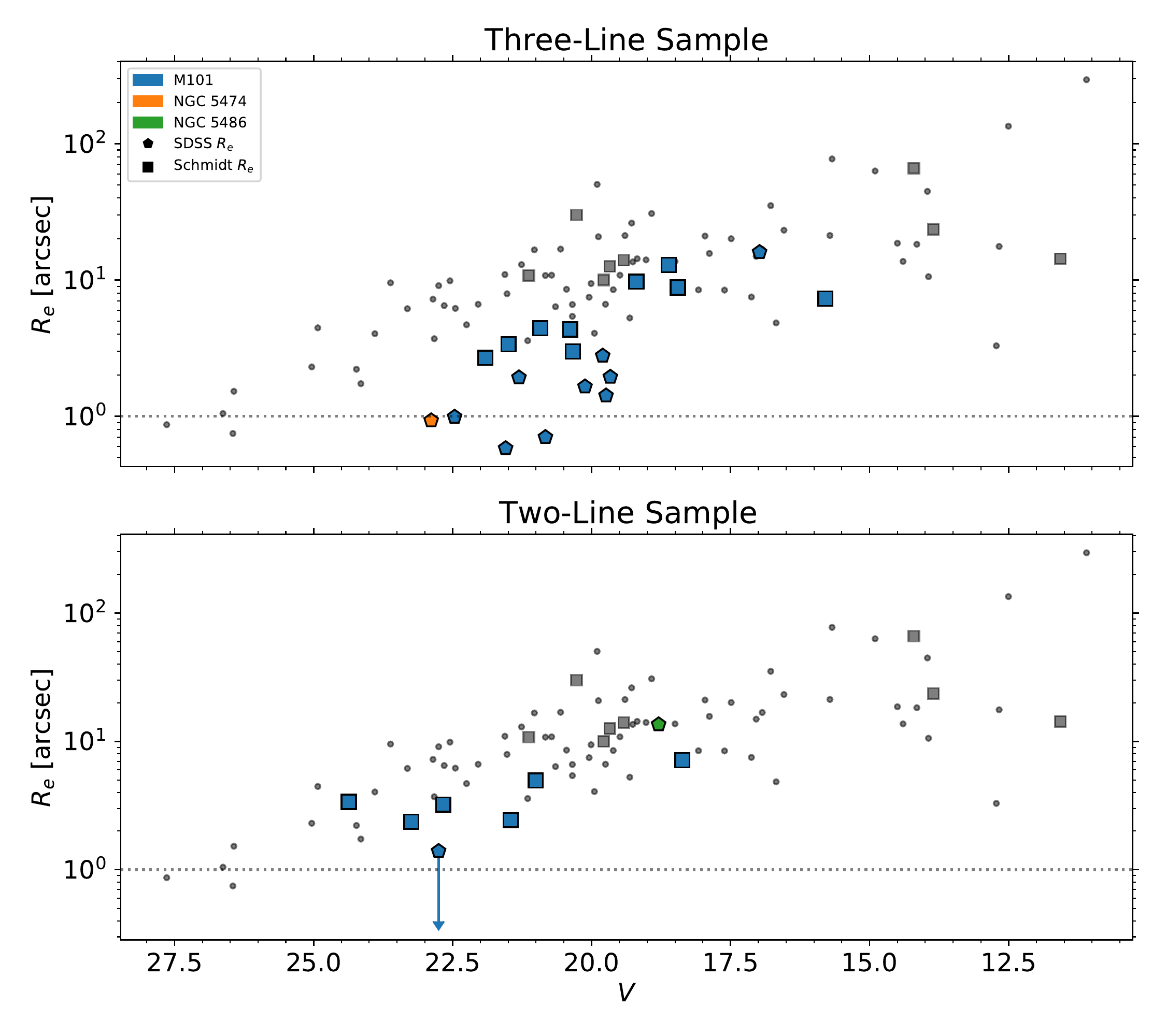}
\caption{Effective radius versus apparent magnitude for our detected sources and comparison samples shifted to the M101 distance. Symbols are the same as in Figure~\ref{struc_all}, but color-coded by associated object: M101 (blue), NGC~5474 (orange), or NGC~5486 (green). Top panel shows the three-line sample; bottom panel shows the two-line sample.\label{struc_homes}}
\end{figure*}

Figure~\ref{struc_homes} shows the three-line and two-line samples' structural information individually, color-coded by associated galaxy. In the three-line sample, most of the sources that appear to be associated with M101 fall along the smaller edge of the trend defined by dwarf galaxies. However, these do not appear to be individual galaxies, but rather \ion{H}{2} regions on the outskirts of the spiral arms (see Figure~\ref{m101_detail}). This includes the two bright sources at $V = 15.8$ and $R_e = \ang{;;7.3}$ and $V = 17$ and $R_e = \ang{;;16}$; they belong to neighboring large \ion{H}{2} complexes directly north of the center of M101 and both are at the end of the large distorted spiral arm. These properties make these \ion{H}{2} regions very similar to the known giant extragalactic \ion{H}{2} regions of M101, such as NGC~5471 \citep{garciabenito2011}.

The source associated with NGC~5474 is the faintest $V$-band source in the three-line sample. It lies just outside $2R_{25}$ and does not appear to be an extension of NGC~5474's spiral arms like the M101-associated sources (see Figure~\ref{dwarf_detail}). Its apparent size is very similar to the Local Group dwarf Segue II, but given that our source is brighter than Segue II by a factor of 30, it is unlikely that our source is an undiscovered, very small dwarf galaxy. Rather, it is likely an \ion{H}{2} region; its physical size is \SI{30}{\parsec}, comparable to the size of a Str\"{o}mgren sphere powered by an O9 star \citep{stromgren1939,osterbrock1989}. 

The two-line sample in Figure~\ref{struc_homes} has a wider variety of sources than the three-line sample. As before, all of the sources associated with M101 appear not as dwarf galaxies, but are interspersed throughout M101's extended spiral arms and are likely just \ion{H}{2} regions in the outer disk. These regions have a larger spread in luminosity than the three-line sources, but are all only a few arcseconds in size. Investigating the line ratios and EWs of the two-line sources explains why these sources are bright enough to be detected in H$\alpha$, but not in H$\beta$: half of the sources have high Balmer decrements, indicating they are heavily extincted, while the other half have such large uncertainties on their H$\beta$ EW that they could reasonably be absorption signatures caused by a strong stellar continuum. 

In the case of the source near NGC~5486, it is unlikely to be a part of the galaxy's outer spiral arms. The galaxy itself has relatively regular outer isophotes in our deep $B$- and $V$-band images (see Figure~\ref{dwarf_detail}). Given that we can resolve the spatially extended nature of the M101 candidates in our images, but cannot for this source, it is unlikely that this is a star-forming object in the M101 Group, but likely more distant. NGC~5486 has a redshift of $z=0.004$ \citep{rc3}; coupled with a Virgocentric flow model of \cite{mould2000}, this gives a distance to the galaxy of \SI{28.2}{\mega\parsec}. If the detected object is at the same distance as NGC~5486, it would have a size of $\sim$\SI{1.8}{\kilo\parsec} and $M_V = -13.5$, similar in physical size to the dwarf irregular galaxy WLM and similar in luminosity to the Sagittarius Dwarf Spheroidal Galaxy  \citep{mcconnachie2012}. 

\subsection{Photometric Analysis}

\begin{figure*}
\epsscale{1.2}
\plotone{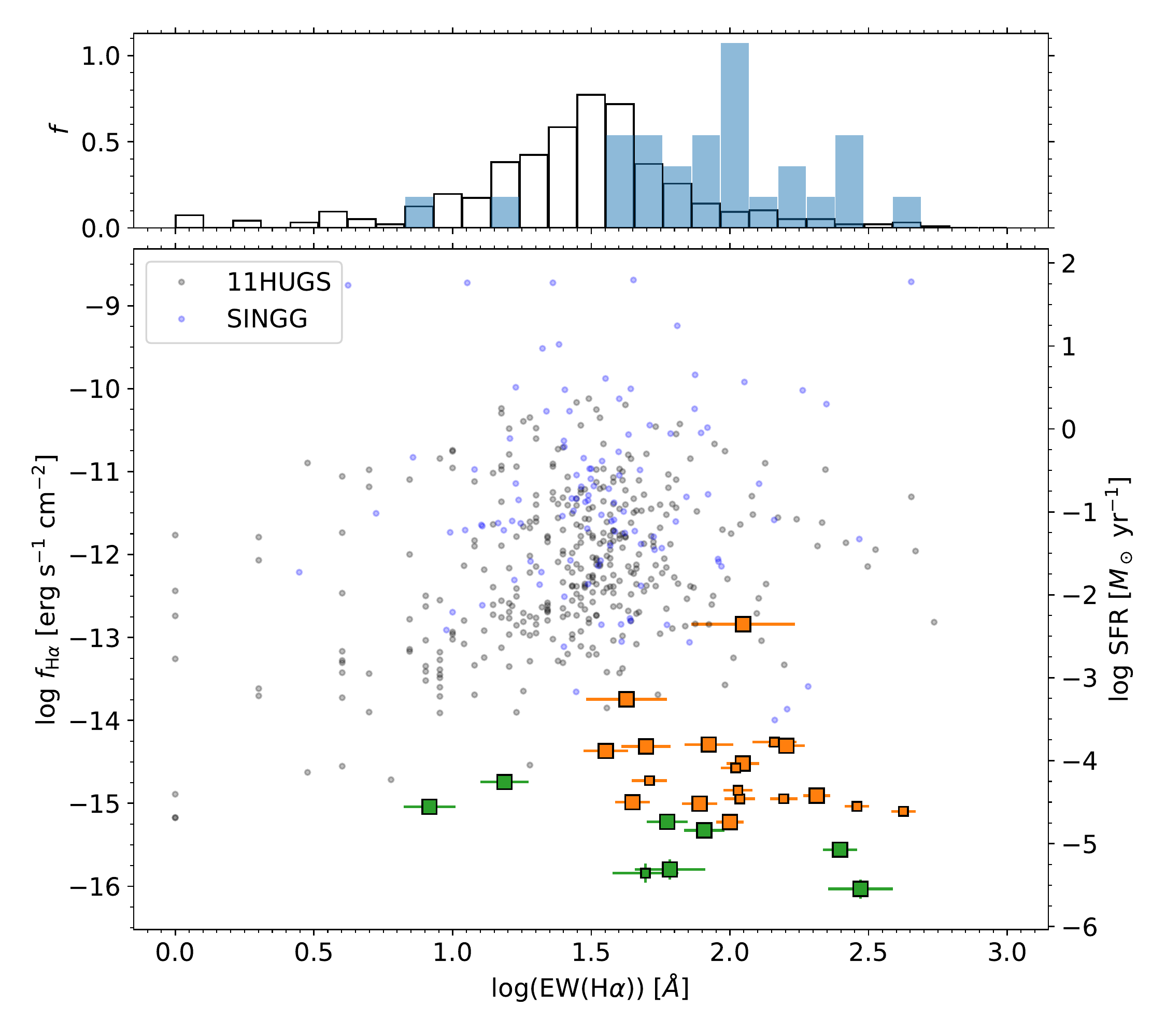}
\caption{The main panel shows H$\alpha$ flux (for our objects) or H$\alpha$-inferred SFR (for galaxies in the 11HUGS and SINGG samples) versus H$\alpha$ equivalent width. Our objects are color-coded by sample and use the left-hand $y$-axis for flux; smaller marker sizes are used for unresolved objects. The 11HUGS (black points) and SINGG (blue points) samples are plotted using the right-hand $y$-axis for absolute SFR. The two axes are equivalent for objects at the \SI{6.9}{\mega\parsec} distance of M101. The upper panel shows the normalized distribution of H$\alpha$ equivalent widths in the various samples, the 11HUGS/SINGG samples together (black outlined bars) and the three- and two-line samples together (light blue bars).\label{ewflux_samples}}
\end{figure*}

Another way of understanding these objects is to compare their star-forming properties to those of known galaxies. We do this by again comparing the observed properties of our samples to the observed properties of known galaxies shifted to the M101 distance. Figure~\ref{ewflux_samples} shows the distribution of H$\alpha$ flux (left axis) and distance-independent equivalent width for the three- and two-line samples. As expected, the three-line sample as a whole has higher EWs than the two-line sample, although both span similar ranges in emission line flux. Other studies have also used EWs to search for extragalactic \ion{H}{2} regions. \cite{werk2010} reported their emission line sources, which they called ELdots, spanned a range of EWs of approximately \SIrange{20}{900}{\angstrom}. Greater than \SI{150}{\angstrom} the majority of their ELdots were H$\alpha$-emitting outlying \ion{H}{2} regions; 7 (\SI{27}{\percent}) of our sources have H$\alpha$ EWs greater than \SI{150}{\angstrom}.

To place our sources in the context of other star-forming galaxies, we also plot in Figure~\ref{ewflux_samples} the H$\alpha$-inferred SFRs (right axis) and EWs for the \SI{11}{\mega\parsec} H$\alpha$ and Ultraviolet Galaxy Survey (11HUGS, \citealt{11hugs,kennicutt2008}) and the Survey for Ionization in Neutral Gas Galaxies (SINGG, \citealt{singg}). The 11HUGS sample is a virtually complete sample of Local Volume galaxies with SFRs in the range of $\sim$\SIrange{e-5}{10}{\solarmass\per\year} and H$\alpha$ EWs spanning \SIrange{0}{545}{\angstrom}. The SINGG sample utilized those \ion{H}{1}-selected galaxies in the \ion{H}{1} Parkes All Sky Survey (HIPASS, \citealt{hipass}); the SINGG sample has SFRs spanning \SIrange{0.0012}{14}{\solarmass\per\year} and H$\alpha$ EWs from \SIrange{2.8}{451}{\angstrom}. 

Comparing our sample to the 11HUGS and SINGG samples, we note several features. On average our sample, if at the M101 distance, would be objects that have SFRs equivalent to the faintest galaxies in the 11HUGS and SINGG samples, i.e.\ less than \SI{e-3}{\solarmass\per\year}. Our sample also consists of objects that have higher H$\alpha$ EWs, on average, than the faint star-forming galaxies in the two galactic samples. This is not surprising given that our survey is an emission line survey; we will find strong emitters with high EWs. If any of our sources are bona-fide dwarfs, they would represent small objects of low star formation rate but high equivalent width, i.e.\ weak starbursting objects. 

There are a few galaxies in the 11HUGS/SINGG samples that have such properties. One galaxy, UGCA~92, a dwarf companion to NGC~1569, has a high EW (\SI{96}{\angstrom}; \citealt{kennicutt2008}) indicating recent star formation. The high EW in UGCA~92 may be due to an interaction with NGC~1569 \citep{makarova2012}, making it an interesting comparison to objects in the M101 Group which may have interacted with M101.

An alternative explanation might be that some of these sources are background objects. In Figure~\ref{ewflux_samples}, there is a population of 11HUGS/SINGG galaxies with high EWs and $-2 \leq \log(\text{SFR}) \leq -1$, three orders of magnitude brighter than our sources would be at the M101 distance. Moving our sources out by a factor of 30 in distance would make them commensurate with those objects in the 11HUGS/SINGG samples. However, at a distance of $\sim$\SI{200}{\mega\parsec}, the emission lines would have been redshifted out of our filters and they would not show up in our detections. Thus it is unlikely that these sources are background objects, and given the mismatch to known star-forming dwarf galaxies as well as the close proximity of many of our sources to M101, the bulk of these sources are likely to be extreme outer-disk \ion{H}{2} regions. 

\begin{figure*}
\epsscale{1.2}
\plotone{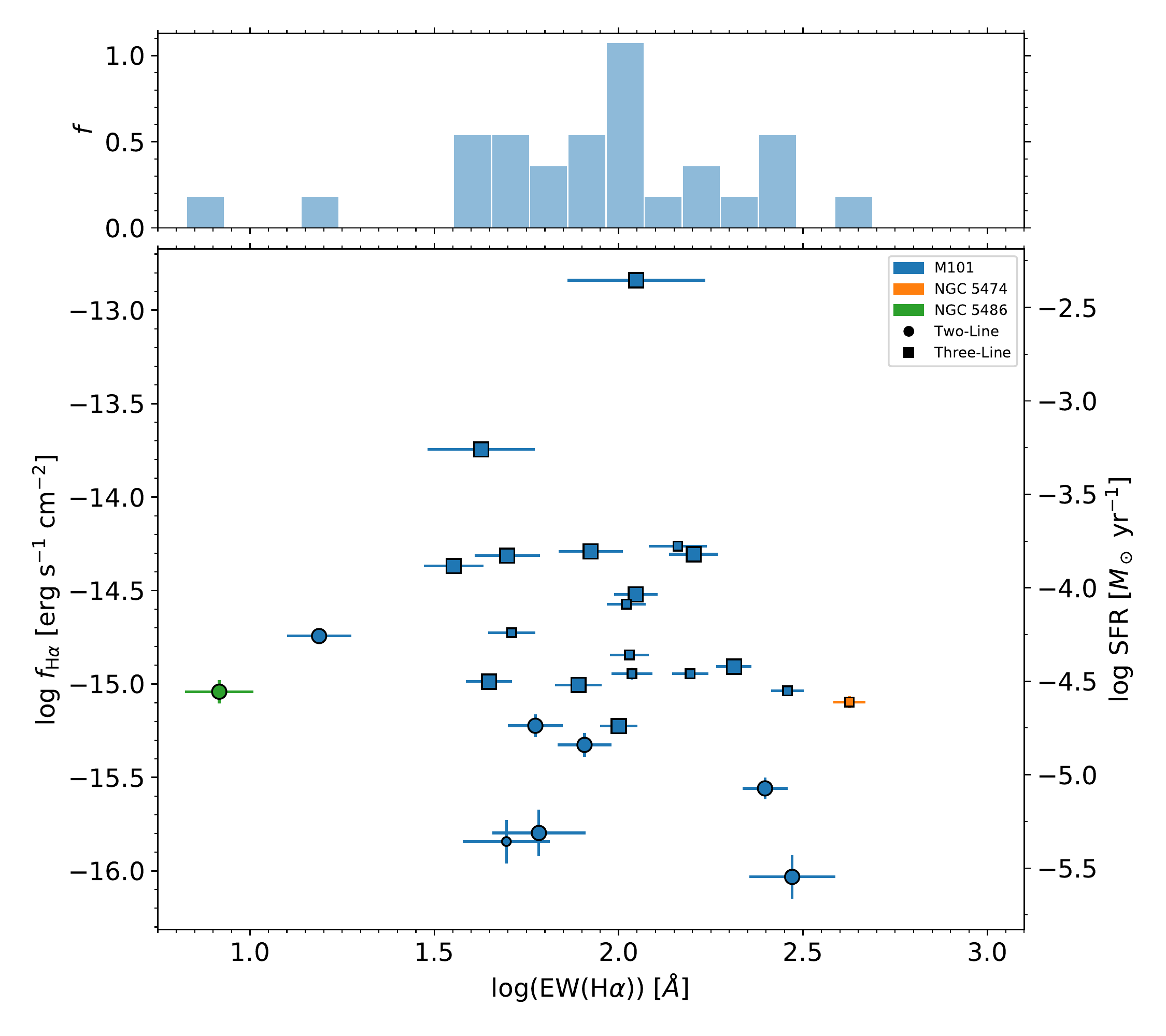}
\caption{The main panel shows H$\alpha$ flux versus EW for points in the three-line sample (squares) and two-line sample (circles), color-coded by associated galaxy. The left $y$-axis shows measured flux; the right $y$-axis shows equivalent star formation if the objects were at the M101 distance. Smaller marker sizes are used for unresolved objects. The upper panel shows the normalized distribution of H$\alpha$ equivalent widths of the sources.\label{ewflux_loc}}
\end{figure*}

Figure~\ref{ewflux_loc} is similar to Figure~\ref{ewflux_samples}, but shows our sources color-coded by their associated object, with the 11HUGS and SINGG samples removed. We notice that the objects with high EWs appear to be associated with M101. This is consistent with the H$\alpha$ EWs of \ion{H}{2} regions in the arms of M101 being high; \cite{cedres2002} found them to range from \SIrange{10}{e4.5}{\angstrom} with a median of \SI{1660}{\angstrom}. This supports our assumption that these sources are indeed associated with M101 as part of its outer spiral structure rather than being outlying \ion{H}{2} regions. 

There are three outliers to the cluster of sources belonging to M101. One source has a very high emission line flux, $f_{\text{H}\alpha} = \SI{1.4e-13}{\erg\per\second\per\square\centi\metre}$, and moderately high H$\alpha$ EW, \SI{112}{\angstrom}. The other has still high but lower values, $f_{\text{H}\alpha} = \SI{1.8e-14}{\erg\per\second\per\square\centi\metre}$ and $\text{EW(H}\alpha) = \SI{42}{\angstrom}$. These are the same outliers in the top panel of Figure~\ref{struc_homes} at $(V, R_e) = (15.8, \ang{;;7.3})$ and $(17,\ang{;;16})$, respectively. The high emission line fluxes supports the hypothesis that these sources are similar to the other giant \ion{H}{2} regions in M101's disk. For instance, the \ion{H}{2} region NGC~5471 has an H$\alpha$ flux of \SI{3.65 \pm 0.17 e-12}{\erg\per\second\per\square\centi\metre} \citep{garciabenito2011}. These sources also have equivalent widths broadly consistent with \ion{H}{2} regions in the inner spiral arms, albeit on the lower end \citep{cedres2002}. 

Conversely, the other outlier falls within the range of emission line fluxes defined by the other M101-associated sources (\SI[multi-part-units=single]{1.8 \pm 0.1 e-15}{\erg\per\second\per\square\centi\metre}), but at a much lower H$\alpha$ EW (\SI[multi-part-units=single]{15.4 \pm 3.1}{\angstrom}). This is the same bright source in the bottom panel of Figure~\ref{struc_homes} at $V = 18.4$ and $R_e = \ang{;;7.1}$. This is located in the same group of \ion{H}{2} complexes as the source described above, but is found in an area of H$\beta$ absorption, indicating a stronger continuum. Given the low EW and H$\beta$ absorption, it is likely that this is a somewhat older \ion{H}{2} region than others that lie in our sample.

The source with the lowest H$\alpha$ EW is the source associated with NGC~5486. If at the distance of NGC 5486, it has an H$\alpha$ luminosity of \SI{8.7e37}{\erg\per\second} giving it an SFR of \SI{4.7e-4}{\solarmass\per\year} \citep{kennicutt2012}. It is interesting to note that its H$\alpha$ luminosity and thus SFR are similar to that of NGC~4163, a nearby dwarf irregular galaxy \citep{kennicutt2008}. Their H$\alpha$ EWs are similar as well; compare our source's \SI{8 \pm 2}{\angstrom} to NGC~4163's \SI{8 \pm 2}{\angstrom} EW \citep{kennicutt2008}. It is quite likely that this source is a star-forming satellite of NGC~5486 similar in structure and luminosity to NGC~4163. 

Finally, we turn to the source associated with NGC~5474 with the highest EW in the entire sample. This source has an H$\alpha$ EW of \SI{422 \pm 43}{\angstrom} putting it firmly in the realm of H$\alpha$-emitting outlying \ion{H}{2} regions as defined by \cite{werk2010}. It also has moderately strong H$\beta$ and [\ion{O}{3}] EWs, \SI{71 \pm 22}{\angstrom} and \SI{282 \pm 37}{\angstrom}, respectively. It has a projected separation from NGC~5474 of $\sim$\ang{;;265} (\SI{8.8}{\kilo\parsec}). Given its structural and photometric properties, this source is likely an outlying \ion{H}{2} region associated with NGC~5474, making it the best (and perhaps only) such candidate discovered in our survey. Followup spectroscopy would be useful to secure its status as a bona-fide object in the M101 Group.

\section{Discussion}

Overall, across the \SI{6}{\square\deg} field of our survey, we detect a total of 19 objects in our three-line sample and 8 objects in our two-line sample. Of these objects, nearly all are found close to M101, aside from one object associated with NGC~5474 and another source associated with NGC~5486 (and thus likely in a background object). None of the emission line sources detected were located far from bright galaxies, arguing against any significant, ongoing intragroup star-forming objects in the M101 Group, down to a limiting star formation rate of \SI{1.7e-6}{\solarmass\per\year}.

In the discussion that follows, we will investigate the consequences of this rarity of star-forming objects in the context of intragroup \ion{H}{1} clouds and newly discovered ultradiffuse galaxies in the M101 Group, as well as the faint end of the star-forming luminosity function. We will also give a more broad discussion of the merger history of M101 and group environments in general. 

Our survey area covers the entire extent of the deepest portion of the M101 \ion{H}{1} imaging survey by \cite{mihos2012}. That survey had a limiting \ion{H}{1} column density of $\log(N) = 16.8$ and \ion{H}{1} mass detection limit of \SI{2e6}{\solarmass}; for comparison, that survey would have detected even the lowest \ion{H}{1} mass objects in the SINGG survey if they were in the M101 Group. The \cite{mihos2012} survey did detect a number of discrete \ion{H}{1} clouds in the M101 Group, along with a diffuse loop of \ion{H}{1} extending \SI{85}{\kilo\parsec} to the southwest of M101. This loop and the associated \ion{H}{1} clouds likely arise from tidal interactions between M101 and its companions, yet our deep narrowband imaging presented here shows no evidence of ongoing star formation in this gas, either in discrete sources or diffuse emission. Nor did the deep broadband imaging of the M101 system by \cite{mihos2013} show evidence for diffuse light in this gas. If extended star formation was triggered in this gas by the past interactions in the M101 Group, that star formation must have been very weak and died out quickly. 

We have also searched for faint H$\alpha$ emission in several of the recently discovered ultradiffuse galaxies in the M101 Group. Five of these dwarfs are seen in our broadband images: DF1, DF2, DF3, DwA, and Dw9 \citep{merritt2014,merritt2016,karachentsev2015,danieli2017,carlsten2019}. \cite{bennet2017} used a large NUV footprint and determined that all five of the dwarfs in our images lack a NUV excess indicating an upper limit SFR of \SI{1.7 \pm 0.5 e-3}{\solarmass\per\year}. We detect no compact emission line sources associated with these objects, and aperature photometry over the \ang{;;10} sizes (typical half-light radii as reported by \citealt{bennet2017}) reveals no diffuse emission down to a level of \SI{4.5e35}{\erg\per\second}. This places a strong upper limit on any star formation in these objects of $\text{SFR} < \SI{2.5e-6}{\solarmass\per\year}$.

This lack of detected H$\alpha$ emission is consistent with the red optical colors of these dwarfs \citep{merritt2014,bennet2017}, arguing that these are older objects, rather than systems formed during recent tidal interactions in the M101 Group. UDGs are observed in both the field and the group environments, so an intuitive formation scenario for those in group environments would be that they were already ``puffed up'' in the field and quenched after falling into the group \citep{roman2017,alabi2018,chan2018,ferremateu2018}. Recent observations seem to support this scenario -- \cite{roman2017} and \cite{alabi2018} found UDGs to have redder colors at smaller cluster-centric distances. Simulations have predicted that of all satellites to have ever existed in a group environment, half originated in the field as UDGs \citep{jiang2019}, themselves formed from supernovae feedback \citep{dicintio2017}, and half were normal galaxies puffed up by tides as satellites \citep{jiang2019}. Perhaps the lopsidedness of M101's satellites might indicate that they are part of a infalling low-mass group \citep{merritt2014}; indeed most isolated blue galaxies, like M101, have lopsided satellite distributions \citep{brainerd2020}. More detailed kinematic investigation is needed of the M101 Group to determine the origin of these UDGs. 

\begin{figure}
\epsscale{1.2}
\plotone{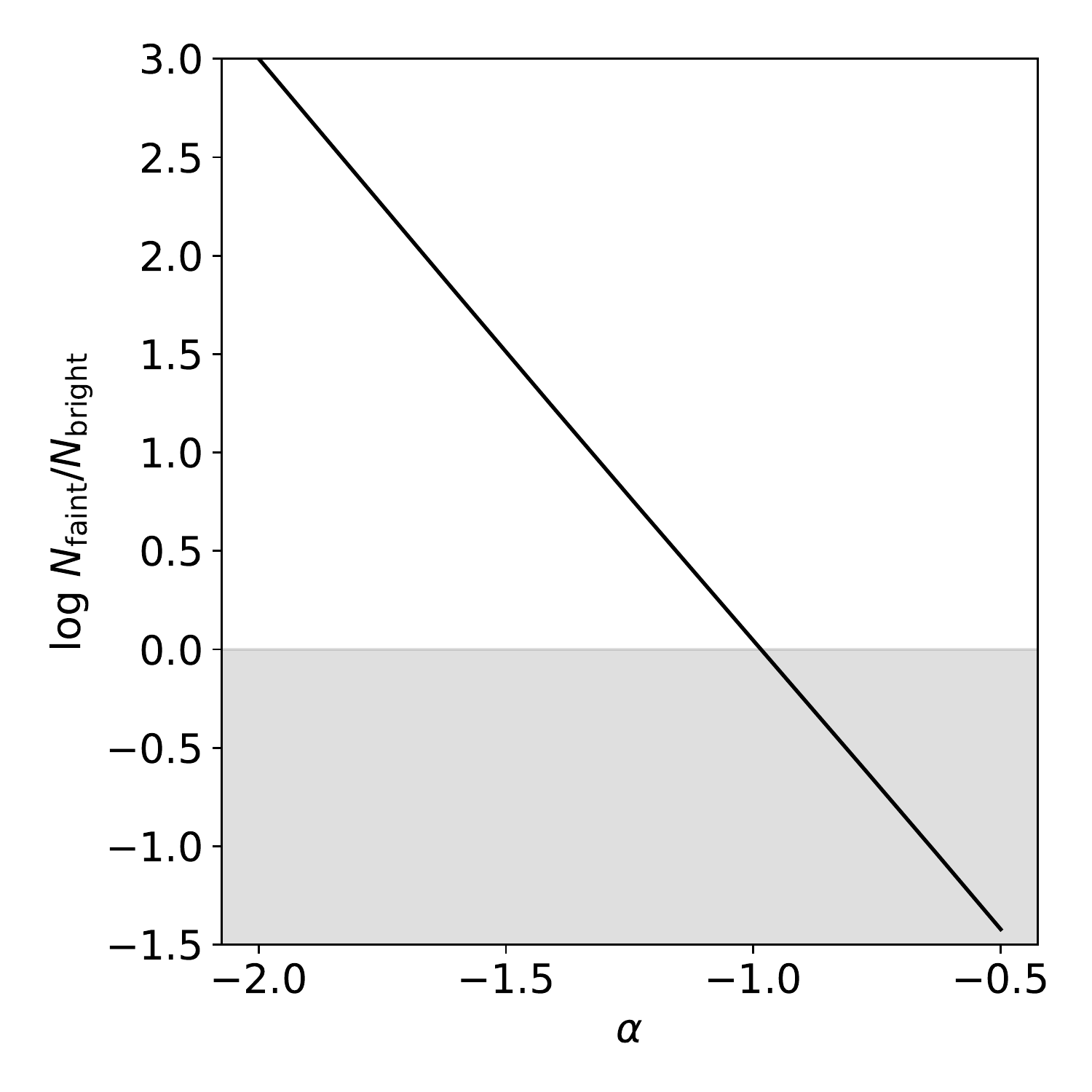}
\caption{The ratio of the expected number of faint-to-bright objects in the M101 Group as a function of the luminosity faint end slope, $\alpha$. Bright objects are defined as having SFRs of $\num{0.01} < \si{\solarmass\per\year} < \num{50}$, while fainter ones have SFRs of $\num{e-5} < \si{\solarmass\per\year} < \num{0.01}$. The gray shaded area shows where the faint sources are rare, as indicated by our data, and favors slopes flatter than $\alpha \sim -1.0$. \label{lf_func}}
\end{figure}

Given the lack of a significant population of low luminosity star-forming dwarf galaxies in the M101 Group, we can place rough limits on the slope of the faint end of the star-forming luminosity function of galaxies in the group. A steep faint end slope would predict a high dwarf-to-giant ratio; the lack of these dwarfs in the M101 Group argues instead for a relatively shallow slope. By adopting a \cite{schechter1976} function with $\text{SFR}_\ast = \SI{9}{\solarmass\per\year}$ \citep{bothwell2011} and varying the faint end slope, we can calculate how many faint star-forming objects we should detect within the group.

In the M101 Group, there are four galaxies with SFRs greater than \SI{0.01}{\solarmass\per\year}: M101, NGC~5474, NGC~5477, and Holmberg~IV \citep{kennicutt2008,kennicutt2012}. However, our deep, wide-field imaging detects no intragroup objects with lower star formation rates. While it is possible that some objects may have been missed due to, for example, being projected directly in front or behind M101's disk, our results argue that the number of faint star forming galaxies (down to $\text{SFRs} \sim \SI{e-5}{\solarmass\per\year}$) must be very low, and certainly do not outnumber the brighter galaxies. This is contrary to expectation if the faint end of the star-forming luminosity function is even moderately steep. For example, as illustrated in Figure~\ref{lf_func}, a flat luminosity function (with slope $\alpha = -1$) would predict equal numbers of objects in the low star formation range of $\num{e-5} < \text{SFR} < \SI{e-2}{\solarmass\per\year}$, while a steeper slope with $\alpha = -1.5$ predicts on the order of 120 objects in the same SFR range. Since we found no evidence of such a population, either flatter values for the faint end slope ($\alpha > -1.0$), or a luminosity function that is sharply truncated below \SI{0.01}{\solarmass\per\year}, are required.

Most of the sources detected (\SI{92}{\percent}) were associated with M101 and are not true outlying \ion{H}{2} regions as they were extensions of the spiral arms. However, one source located near NGC~5474 had both photometric and structural properties consistent with previous searches for isolated \ion{H}{2} regions. We can quantify this source by calculating the total number of ionizing photons, $Q_0$, from \cite{osterbrock1989}:
\begin{equation}
Q_0 = 2.2\frac{L_{\text{H}\alpha}}{h\nu_{\text{H}\alpha}} \sim \num{7.2e11}L_{\text{H}\alpha} \, \si{\per\second},
\end{equation}
where we assume case B recombination \citep{osterbrock1989}. Assuming this source to be at the M101 distance, it could be powered by only 4 O9V stars \citep{martins2005}, similar to the faintest isolated \ion{H}{2} regions detected in the studies of other galaxies by \cite{ryan-weber2004} and \cite{werk2010}. Such a low level of star formation pushes the ill-defined boundary differentiating outlying \ion{H}{2} regions from star-forming dwarf galaxies; follow-up spectroscopy would be of interest in studying this source in more detail. 

Another source that deserves spectroscopic follow-up is N5486-2-1. Although this was located within $2R_{25}$ of NGC~5486 and does not satisfy the classical definition of an outlying \ion{H}{2} region, assuming it is physically near NGC~5486, it displays many properties similar to local dwarf galaxies. The knot of star formation our survey targeted could be powered by $\sim$80 O9V stars \citep{martins2005}, consistent with dwarf galaxies \citep{werk2010}. Additionally, this source has a similar $V$-band and H$\alpha$ luminosity to the dwarf irregular galaxy NGC~4163 as mentioned above. Another striking resemblance is the asymmetry in the broadband images. The source appears to consist of a bright compact region with a more diffuse extension to the west, structure which is also evident in the shallower $gri$ composite from SDSS. Similarly NGC~4163 is itself asymmetric in nature, with a burst of star formation occurring in the central portion of the galaxy, but not the outer regions \citep{mcquinn2012}. It also has a peculiar \ion{H}{1} distribution, with an \ion{H}{1} tail to the west and possibly the south \citep{hunter2011,lelli2014}. Although it is not clear what could be causing the behavior in NGC~4163, perhaps some large-scale interaction with NGC~5486 has caused this asymmetry in our source.

Equally uncertain is the interaction history of the M101 Group as a whole. The asymmetric disk of M101 has long been believed to arise from an interaction \citep{beale1969,rownd1994,waller1997}. Low surface brightness optical light has been detected in the outskirts of M101 \citep{mihos2013} with colors and stellar populations consistent with a burst of star formation $\sim$\SIrange{300}{400}{\mega\year} ago \citep{mihos2018}. High-velocity \ion{H}{1} gas has also been observed in the same location as the optical light \citep{vanderhulst1988,mihos2012}, while intermediate-velocity \ion{H}{1} gas has been detected between M101 and NGC~5474 \citep{mihos2012}. NGC~5474's offset bulge is usually added as further evidence of an interaction, although recent work has called that into question \citep{bellazzini2020,pascale2021}. 

Galaxy-galaxy interactions frequently give rise to star formation in tidal debris \citep{schombert1990,gerhard2002,sakai2002,cortese2004,ryan-weber2004,boquien2007,boquien2009,werk2010}. Intragroup \ion{H}{2} regions have also been detected between pairs of massive galaxies in compact groups \citep{sakai2002,oliveira2004}, illustrating that the small group environment easily drives interactions leading to star-forming regions beyond the galactic disks.

Given our lack of detection of any isolated, intragroup \ion{H}{2} regions, what does this mean for the interaction scenario? All of the examples of star-forming objects in interacting systems involve strong interactions or major-merger events. The lack of star-forming \ion{H}{2} regions between NGC~5474 and M101 might indicate that the two galaxies have undergone only a weak interaction, or that any tidally-triggered intragroup star formation was very short-lived. If the pair's luminosity ratio of $17:1$ in the $V$-band is indicative of the masses of the galaxies, then this would classify it as a low-mass encounter. 

Furthermore, the M101 Group is not a compact galaxy group. Compact galaxy groups are very dense environments, containing only a few galaxies separated by distances comparable to their sizes. This makes them strongly interacting environments, leading to the formation of tidal tails, intragroup star formation, and tidal dwarf galaxies \citep[e.g.][]{demello2008,torresflores2009}. A well-known example of this is Stephan's Quintet (Arp 319) where the star formation is heavily influenced by the ongoing interactions between its member galaxies. Intragroup star formation can be found at the tip of a shock front \citep{xu1999,xu2003} and in a tidal tail \citep{arp1973,sulentic2001,xu2005}. Numerous tidal dwarf galaxy candidates have been found in tidal tails as well \citep{hunsberger1996}. 

In contrast with Stephan's Quintet, the M101 Group is dominated by M101, with relatively few low-mass companions, making it possibly the poorest group in the Local Volume \citep{bremnes1999}. The lack of an abundance of intragroup star formation in the M101 Group might be because the M101 Group involves only weak interactions with low mass companions. Given M101's ``anemic'' stellar halo \citep{vandokkum2014,jang2020}, it is unlikely that M101 has gone through any major mergers, and very few, if any, minor mergers. With no comparable companions nearby, it is likely that no outlying \ion{H}{2} regions will form until one of the low mass companions falls into and merges with M101. Clearly, detailed computer modeling is needed to unravel all of the features of the M101 Group. 

\section{Summary}

We have conducted a narrowband emission line survey of the M101 Group to search for faint star-forming dwarf galaxies and outlying \ion{H}{2} regions. Using narrowband filters that target H$\alpha$, H$\beta$, and [\ion{O}{3}], our survey detects sources as faint as $f_{\text{H}\alpha} = \SI{5.7e-17}{\erg\per\second\per\square\centi\metre}$, or an equivalent star formation rate of \SI{1.7e-6}{\solarmass\per\year} for sources in the M101 Group. Imaging in multiple lines significantly reduces contamination from single-line detections of redshifted sources in the background, but there is significant contamination from Milky Way stars, particularly M stars whose complicated continuum structure mimics emission line flux in our narrowband filters. To eliminate this contamination, we use stellar spectral templates to place equivalent width cuts on the sample, requiring $\text{EW}(\text{H}\alpha) > \SI{8}{\angstrom}$, $\text{EW}(\text{H}\beta) > \SI{2}{\angstrom}$, and $\text{EW}(\text{[\ion{O}{3}]}) > \SI{5}{\angstrom}$. We further cross-match the sample with the \emph{Gaia} EDR3 catalog to eliminate objects with detected parallax and proper motion. Finally, we cross-match the remaining sources with the SDSS catalog to remove known background sources and, in cases where H$\beta$ is detected, we also reject sources with anomalous (unphysical) Balmer decrements. Our final sample consists of 17 sources detected in all three of the H$\alpha$, H$\beta$, and [\ion{O}{3}] filters (the three-line sample), and another 8 sources detected only in H$\alpha$ and [\ion{O}{3}] (the two-line sample). 

\begin{enumerate}
    \item Of the 27 total sources, 25 (\SI{93}{\percent}) were located in M101's extreme outer disk, while two were associated with other bright galaxies in the survey area. We found no evidence for any truly outlying emission line regions in the M101 intragroup environment.
    
    \item Assuming that our sources are at the M101 distance, a comparison between their physical properties and the properties of known Local Group dwarfs and star-forming galaxies in the 11HUGS and SINGG surveys argues that most of our sources are not bona-fide star-forming dwarfs. Instead, their small sizes, low H$\alpha$ luminosities, and high H$\alpha$ EWs argue that they are more likely M101 outer disk \ion{H}{2} regions.
    
    \item Two detected sources were unassociated with M101 itself: an unresolved but high EW source near the M101 satellite NGC~5474 (likely an outlying \ion{H}{2} region), and a spatially extended, low EW satellite of the background galaxy NGC 5486.
    
    \item We searched for ionized gas emission in five previously discovered ultradiffuse dwarf galaxy candidates in the M101 Group (DF1, DF2, DF3, DwA, Dw9) and found none to a limit of \SI{2.5e-6}{\solarmass\per\year}. This makes it unlikely that any of these ultradiffuse candidates are undergoing star formation at present times. 
    
    \item The lack of any significant population of star-forming objects in the M101 Group down to a limiting equivalent star formation rate of \SI{e-5}{\solarmass\per\year} argues for a shallow faint end slope of the star-forming luminosity function ($\alpha \sim 1$).
    
    \item Given that tidally induced star formation is a common outcome of close galaxy encounters, the lack of outlying \ion{H}{2} regions also suggests either that interactions within the M101 Group have been relatively weak, or that any extended star formation triggered by the encounters has quickly died out. Both scenarios are consistent with recent studies attributing M101's distorted morphology to a weak, retrograde encounter with NGC~5474 some \SIrange{300}{400}{\mega\year} ago. 
    
\end{enumerate}

\begin{acknowledgments}

This work was supported in part by a Towson Memorial Scholarship to RG, and by grants to JCM from the National Science Foundation (award 1108964) and the Mt Cuba Astronomical Foundation. We also thank Stacy McGaugh for helpful discussions.

\end{acknowledgments}

\facilities{CWRU:Schmidt}

\software{Astropy \citep{astropy2013,astropy2018}, Matplotlib \citep{matplotlib}, NumPy \citep{numpy}}

\movetabledown=0.7in
\begin{longrotatetable}
\begin{deluxetable*}{lccccccccccc}
\tablecaption{Narrowband Properties\label{narrow_gold}}
\tablehead{
\colhead{ID} & \colhead{RA} & \colhead{Dec} & \colhead{E/U} & \colhead{$\Delta f_{\text{H}\alpha}$\tablenotemark{a}} & \colhead{$\Delta f_{\text{H}\beta}$} & \colhead{$\Delta f_{\text{[\ion{O}{3}]}}$} & \colhead{EW$_{\text{H}\alpha}$\tablenotemark{a}} & \colhead{EW$_{\text{H}\beta}$} & \colhead{EW$_{\text{[\ion{O}{3}]}}$} & \colhead{$\text{H}\alpha/\text{H}\beta$\tablenotemark{a}} & \colhead{$\text{[\ion{O}{3}]}/\text{H}\alpha$\tablenotemark{a}}  \\
\colhead{} & \colhead{{[Deg]}} & \colhead{{[Deg]}} & \colhead{} & \colhead{{[\SI{e-16}{\erg\per\second\per\square\centi\metre}]}}  & \colhead{{[\SI{e-16}{\erg\per\second\per\square\centi\metre}]}} & \colhead{{[\SI{e-16}{\erg\per\second\per\square\centi\metre}]}} & \colhead{{[\si{\angstrom}]}} & \colhead{{[\si{\angstrom}]}} & \colhead{{[\si{\angstrom}]}} & \colhead{} & \colhead{} }
\startdata
M101-3-1 & 210.7923 & 54.5976 & E & 51.167 (1.637) & 19.265 (2.275) & 20.744 (2.190) & 83.947 (16.911) & 22.162 (6.247) & 26.814 (6.174) & 2.656 & 0.405 \\ 
M101-3-2 & 210.8047 & 54.5948 & E & 1446.379 (3.644) & 520.204 (4.936) & 1827.889 (4.853) & 111.643 (48.094) & 25.178 (13.898) & 102.963 (45.658) & 2.780 & 1.264 \\ 
M101-3-3 & 210.8159 & 54.5954 & E & 42.782 (1.494) & 8.897 (2.075) & 19.030 (2.001) & 35.695 (6.600) & 4.523 (1.493) & 11.347 (2.435) & 4.809 & 0.445 \\ 
M101-3-4 & 210.8268 & 54.5884 & E & 179.646 (2.756) & 28.383 (3.829) & 172.745 (3.704) & 42.363 (14.186) & 3.854 (1.740) & 26.896 (9.307) & 6.329 & 0.962 \\ 
M101-3-5 & 210.8277 & 54.5944 & E & 48.542 (1.654) & 13.523 (2.300) & 40.987 (2.218) & 49.854 (10.167) & 8.679 (2.687) & 30.387 (6.503) & 3.590 & 0.844 \\ 
M101-3-6 & 210.9173 & 54.6299 & U & 9.192 (0.629) & 4.339 (0.874) & 8.228 (0.842) & 287.044 (29.440) & 254.578 (57.042) & 511.041 (65.872) & 2.118 & 0.895 \\ 
M101-3-7 & 211.0488 & 54.5866 & U & 14.315 (0.858) & 6.166 (1.192) & 9.960 (1.148) & 106.914 (12.851) & 37.397 (8.795) & 63.710 (10.026) & 2.322 & 0.696 \\ 
M101-3-8 & 211.0629 & 54.5992 & E & 5.951 (0.558) & 3.063 (0.774) & 2.935 (0.745) & 100.021 (11.582) & 87.032 (23.253) & 61.645 (16.232) & 1.943 & 0.493 \\ 
M101-3-9 & 211.1460 & 54.4876 & U & 11.338 (0.848) & 6.481 (1.178) & 6.802 (1.133) & 108.648 (13.840) & 73.776 (16.600) & 87.410 (17.255) & 1.749 & 0.600 \\ 
M101-3-10 & 211.1545 & 54.4514 & U & 18.797 (1.105) & 5.446 (1.537) & 12.977 (1.480) & 51.262 (7.518) & 9.571 (3.168) & 28.209 (5.053) & 3.452 & 0.690 \\ 
M101-3-11 & 211.1651 & 54.4407 & U & 26.721 (0.961) & 10.942 (1.331) & 31.493 (1.285) & 104.887 (12.783) & 32.559 (6.279) & 113.648 (14.358) & 2.442 & 1.179 \\ 
M101-3-12 & 211.1661 & 54.4580 & U & 54.601 (1.483) & 22.772 (2.056) & 107.590 (1.990) & 144.730 (26.330) & 58.293 (14.470) & 360.163 (66.812) & 2.398 & 1.970 \\ 
M101-3-13 & 211.1676 & 54.2532 & E & 49.412 (1.249) & 22.497 (1.728) & 112.280 (1.678) & 159.882 (24.508) & 76.696 (16.012) & 426.537 (66.469) & 2.196 & 2.272 \\ 
M101-3-14 & 211.1932 & 54.4407 & E & 10.295 (0.929) & 4.341 (1.290) & 7.437 (1.243) & 44.510 (6.442) & 17.032 (5.635) & 31.805 (6.473) & 2.372 & 0.722 \\ 
M101-3-15 & 211.1955 & 54.3812 & E & 9.888 (0.917) & 4.552 (1.278) & 4.716 (1.228) & 77.771 (11.291) & 28.326 (8.932) & 36.478 (10.382) & 2.172 & 0.477 \\ 
M101-3-16 & 211.1994 & 54.4902 & U & 11.352 (0.755) & 5.820 (1.050) & 3.253 (1.009) & 156.237 (17.710) & 79.776 (17.197) & 49.120 (15.925) & 1.950 & 0.287 \\ 
M101-3-17 & 211.2051 & 54.3800 & E & 30.155 (1.081) & 13.725 (1.498) & 9.856 (1.439) & 111.209 (15.120) & 49.666 (9.963) & 46.681 (9.261) & 2.197 & 0.327 \\ 
M101-3-18 & 211.2138 & 54.3906 & E & 12.376 (0.755) & 5.801 (1.049) & 11.343 (1.011) & 205.514 (22.631) & 90.213 (19.485) & 204.745 (26.554) & 2.133 & 0.917 \\ 
N5474-3-1 & 211.2695 & 53.7356 & U & 8.000 (0.578) & 2.646 (0.804) & 7.130 (0.774) & 422.339 (42.488) & 70.703 (22.404) & 281.790 (36.729) & 3.023 & 0.891 \\ 
\enddata
\tablenotetext{a}{H$\alpha$ fluxes and EWs have been corrected for [\ion{N}{2}] emission by assuming [\ion{N}{2}]/H$\alpha = 0.33$ \citep{kennicutt1992,jansen2000}.}
\tablecomments{Numbers in parentheses are the associated uncertainties for that quantity.}
\end{deluxetable*}
\end{longrotatetable}

\movetabledown=0.7in
\begin{longrotatetable}
\begin{deluxetable*}{lccccccccccc}
\tablecaption{Narrowband Properties\label{narrow_silver}}
\tablehead{
\colhead{ID} & \colhead{RA} & \colhead{Dec} & \colhead{E/U} & \colhead{$\Delta f_{\text{H}\alpha}$\tablenotemark{a}} & \colhead{$\Delta f_{\text{H}\beta}$} & \colhead{$\Delta f_{\text{[\ion{O}{3}]}}$} & \colhead{EW$_{\text{H}\alpha}$\tablenotemark{a}} & \colhead{EW$_{\text{H}\beta}$} & \colhead{EW$_{\text{[\ion{O}{3}]}}$} & \colhead{$\text{H}\alpha/\text{H}\beta$\tablenotemark{a}} & \colhead{$\text{[\ion{O}{3}]}/\text{H}\alpha$\tablenotemark{a}}  \\
\colhead{} & \colhead{{[Deg]}} & \colhead{{[Deg]}} & \colhead{} & \colhead{{[\SI{e-16}{\erg\per\second\per\square\centi\metre}]}}  & \colhead{{[\SI{e-16}{\erg\per\second\per\square\centi\metre}]}} & \colhead{{[\SI{e-16}{\erg\per\second\per\square\centi\metre}]}} & \colhead{{[\si{\angstrom}]}} & \colhead{{[\si{\angstrom}]}} & \colhead{{[\si{\angstrom}]}} & \colhead{} & \colhead{} }
\startdata
M101-2-1 & 210.8156 & 54.5862 & E & 18.049 (1.500) & -9.414 (2.092) & 29.587 (2.025) & 15.397 (3.094) & -4.529 (-1.467) & 15.829 (3.183) & -1.917 & 1.639 \\ 
M101-2-2 & 210.9127 & 54.5898 & E & 1.596 (0.457) & 0.374 (0.638) & 3.050 (0.614) & 60.755 (17.741) & 11.030 (18.830) & 125.374 (26.252) & 4.267 & 1.911 \\ 
M101-2-3 & 211.1586 & 54.2476 & E & 5.978 (0.824) & 1.448 (1.150) & 6.867 (1.107) & 59.436 (10.137) & 8.661 (6.968) & 48.588 (9.309) & 4.128 & 1.149 \\ 
M101-2-4 & 211.1632 & 54.4533 & E & 4.724 (0.684) & 0.524 (0.955) & 5.957 (0.919) & 80.791 (13.505) & 5.584 (10.199) & 79.426 (14.028) & 9.022 & 1.261 \\ 
M101-2-5 & 211.1730 & 54.4580 & E & 0.929 (0.248) & 0.347 (0.345) & 1.049 (0.333) & 295.520 (79.335) & 117.886 (117.421) & 325.623 (103.728) & 2.679 & 1.129 \\ 
M101-2-6 & 211.2291 & 54.3049 & U & 1.436 (0.386) & 0.570 (0.537) & 1.782 (0.517) & 49.596 (13.533) & 17.634 (16.640) & 70.720 (20.794) & 2.518 & 1.241 \\ 
M101-2-7 & 211.2296 & 54.2999 & E & 2.764 (0.369) & 1.219 (0.513) & 3.219 (0.494) & 249.438 (35.095) & 97.326 (41.319) & 265.051 (42.508) & 2.267 & 1.165 \\ 
N5486-2-1 & 211.8695 & 55.0837 & E & 9.093 (1.306) & 3.388 (1.800) & 8.630 (1.741) & 8.253 (1.771) & 3.254 (1.849) & 7.853 (2.036) & 2.683 & 0.949 \\ 
\enddata
\tablenotetext{a}{H$\alpha$ fluxes and EWs have been corrected for [\ion{N}{2}] emission by assuming [\ion{N}{2}]/H$\alpha = 0.33$ \citep{kennicutt1992,jansen2000}.}
\tablecomments{Numbers in parentheses are the associated uncertainties for that quantity.}
\end{deluxetable*}
\end{longrotatetable}

\begin{deluxetable*}{lccccccccccc}
\tablecaption{Three-Line Sample Broadband Properties\label{broad_gold}}
\tablehead{
\colhead{ID} & \colhead{$V$} & \colhead{$B-V$} & \colhead{$u$} & \colhead{$g$} & \colhead{$r$} & \colhead{$i$} & \colhead{$z$} & \colhead{$FUV$} & \colhead{$NUV$}}
\startdata
M101-3-1 & 19.20 & 0.05 & -- & -- & -- & -- & -- & -- & -- \\ 
M101-3-2 & 15.80 & 0.10 & 21.50 & 21.60 & 21.39 & 21.10 & 20.55 & -- & -- \\ 
M101-3-3 & 18.45 & 0.05 & -- & -- & -- & -- & -- & -- & -- \\ 
M101-3-4 & 16.98 & 0.09 & 18.71 & 18.56 & 18.65 & 18.92 & 22.70 & -- & -- \\ 
M101-3-5 & 18.61 & 0.10 & -- & -- & -- & -- & -- & -- & -- \\ 
M101-3-6 & 22.46 & 0.28 & 23.67 & 22.49 & 23.36 & 23.79 & 23.54 & -- & -- \\ 
M101-3-7 & 20.83 & 0.06 & 22.47 & 22.33 & 22.23 & 24.29 & 23.24 & -- & -- \\ 
M101-3-8 & 21.91 & 0.32 & -- & -- & -- & -- & -- & -- & -- \\ 
M101-3-9 & 21.31 & -0.01 & 21.33 & 21.50 & 21.64 & 22.19 & 22.61 & 17.55 & 18.87 \\ 
M101-3-10 & 19.74 & 0.05 & 20.72 & 19.93 & 20.12 & 21.25 & 21.36 & 16.99 & 18.14 \\ 
M101-3-11 & 20.12 & 0.12 & 21.07 & 21.13 & 21.23 & 22.02 & 22.18 & 17.23 & 18.40 \\ 
M101-3-12 & 19.66 & 0.27 & 21.57 & 20.73 & 20.66 & 22.35 & 21.47 & 16.28 & 17.47 \\ 
M101-3-13 & 19.80 & 0.59 & 19.79 & 19.71 & 19.78 & 20.81 & 19.24 & 16.48 & 17.96 \\ 
M101-3-14 & 20.34 & 0.41 & -- & -- & -- & -- & -- & 17.05 & 18.34 \\ 
M101-3-15 & 20.92 & 0.10 & -- & -- & -- & -- & -- & 16.96 & 18.35 \\ 
M101-3-16 & 21.55 & 0.15 & 22.46 & 22.14 & 23.03 & 22.84 & 21.40 & 17.49 & 18.73 \\ 
M101-3-17 & 20.39 & 0.08 & -- & -- & -- & -- & -- & 16.56 & 17.94 \\ 
M101-3-18 & 21.49 & 0.26 & -- & -- & -- & -- & -- & 17.30 & 18.69 \\ 
N5474-3-1 & 22.89 & -0.16 & 22.83 & 22.46 & 22.77 & 23.74 & 23.61 & -- & -- \\ 
\enddata
\end{deluxetable*}

\begin{deluxetable*}{lccccccccccc}
\tablecaption{Two-Line Sample Broadband Properties\label{broad_silver}}
\tablehead{
\colhead{ID} & \colhead{$V$} & \colhead{$B-V$} & \colhead{$u$} & \colhead{$g$} & \colhead{$r$} & \colhead{$i$} & \colhead{$z$} & \colhead{$FUV$} & \colhead{$NUV$}}
\startdata
M101-2-1 & 18.37 & 0.08 & -- & -- & -- & -- & -- & -- & -- \\ 
M101-2-2 & 22.67 & 0.07 & -- & -- & -- & -- & -- & -- & -- \\ 
M101-2-3 & 21.01 & 0.10 & -- & -- & -- & -- & -- & 17.39 & 18.86 \\ 
M101-2-4 & 21.45 & 0.21 & -- & -- & -- & -- & -- & 17.98 & 19.22 \\ 
M101-2-5 & 24.37 & 0.25 & -- & -- & -- & -- & -- & 20.12 & 21.36 \\ 
M101-2-6 & 22.75 & 0.35 & 23.35 & 22.56 & 22.34 & 22.44 & 22.42 & 18.73 & 20.28 \\ 
M101-2-7 & 23.24 & 0.31 & -- & -- & -- & -- & -- & 18.83 & 20.38 \\ 
N5486-2-1 & 18.79 & 0.55 & 19.51 & 19.17 & 19.24 & 19.14 & 23.58 & -- & -- \\ 
\enddata
\end{deluxetable*}

\bibliography{Bibliography.bib}
\bibliographystyle{aasjournal.bst}

\end{document}